\documentclass[sigconf]{acmart} 

\AtBeginDocument{%
  \providecommand\BibTeX{{%
    \normalfont B\kern-0.5em{\scshape i\kern-0.25em b}\kern-0.8em\TeX}}}

\copyrightyear{2026}
\acmYear{2026}
\setcopyright{cc}
\setcctype{by}
\acmConference[IUI '26]{31st International Conference on Intelligent User Interfaces}{March 23--26, 2026}{Paphos, Cyprus}
\acmBooktitle{31st International Conference on Intelligent User Interfaces (IUI '26), March 23--26, 2026, Paphos, Cyprus}
\acmDOI{10.1145/3742413.3789099}
\acmISBN{979-8-4007-1984-4/26/03}

\usepackage{subcaption}
\usepackage{multirow}
\usepackage{xspace}
\usepackage{fontawesome}
\usepackage{enumitem}
\usepackage{fancyvrb}
\usepackage{listings}

\newcommand{\projectname}{CoLyricist}

\begin{document}

\title{CoLyricist: Enhancing Lyric Writing with AI through Workflow-Aligned Support}

\author{Masahiro Yoshida}
\authornote{Both authors contributed equally to this research.}
\affiliation{%
  \institution{University of California, Los Angeles}
  \city{Los Angeles}
  \state{California}
  \country{USA}
}
\email{masahiro.yoshida.research@gmail.com}
\orcid{0009-0000-2394-1725}

\author{Bingxuan Li}
\authornotemark[1]
\affiliation{%
  \institution{University of California, Los Angeles}
  \city{Los Angeles}
  \state{California}
  \country{USA}
}
\email{bingxuan@ucla.edu}
\orcid{0000-0002-9193-5308}

\author{Songyan Zhao}
\affiliation{%
  \institution{University of California, Los Angeles}
  \city{Los Angeles}
  \state{California}
  \country{USA}
}
\email{zhaosongyan@g.ucla.edu}
\orcid{0009-0000-6116-5325}

\author{Qinyi Zhou}
\affiliation{%
  \institution{University of California, Los Angeles}
  \city{Los Angeles}
  \state{California}
  \country{USA}
}
\email{qinyizhou24@g.ucla.edu}
\orcid{0009-0000-1878-3646}

\author{Shiwei Hu}
\affiliation{%
  \institution{University of California, Los Angeles}
  \city{Los Angeles}
  \state{California}
  \country{USA}
}
\email{shiweihu@ucla.edu}
\orcid{0009-0003-1662-4846}

\author{Xiang 'Anthony' Chen}
\affiliation{%
  \institution{University of California, Los Angeles}
  \city{Los Angeles}
  \state{California}
  \country{USA}
}
\email{xac@ucla.edu}
\orcid{0000-0002-8527-1744}

\author{Nanyun Peng}
\affiliation{%
  \institution{University of California, Los Angeles}
  \city{Los Angeles}
  \state{California}
  \country{USA}
}
\email{violetpeng@cs.ucla.edu}
\orcid{0000-0002-8509-6595}

\renewcommand{\shortauthors}{Yoshida and Li, et al.}

\begin{abstract}
  We propose CoLyricist, an AI-assisted lyric writing tool designed to support the typical workflows of experienced lyricists and enhance their creative efficiency. While lyricists have unique processes, many follow common stages. Tools that fail to accommodate these stages challenge integration into creative practices. Existing research and tools lack sufficient understanding of these songwriting stages and their associated challenges, resulting in ineffective designs. Through a formative study involving semi-structured interviews with 10 experienced lyricists, we identified four key stages: Theme Setting, Ideation, Drafting Lyrics, and Melody Fitting. CoLyricist addresses these needs by incorporating tailored AI-driven support for each stage, optimizing the lyric writing process to be more seamless and efficient. To examine whether this workflow-aligned design also benefits those without prior experience, we conducted a user study with 16 participants, including both experienced and novice lyricists. Results showed that CoLyricist enhances the songwriting experience across skill levels. Novice users especially appreciated the Melody-Fitting feature, while experienced users valued the Ideation support.

\end{abstract}




\begin{teaserfigure}
  \includegraphics[width=\textwidth]{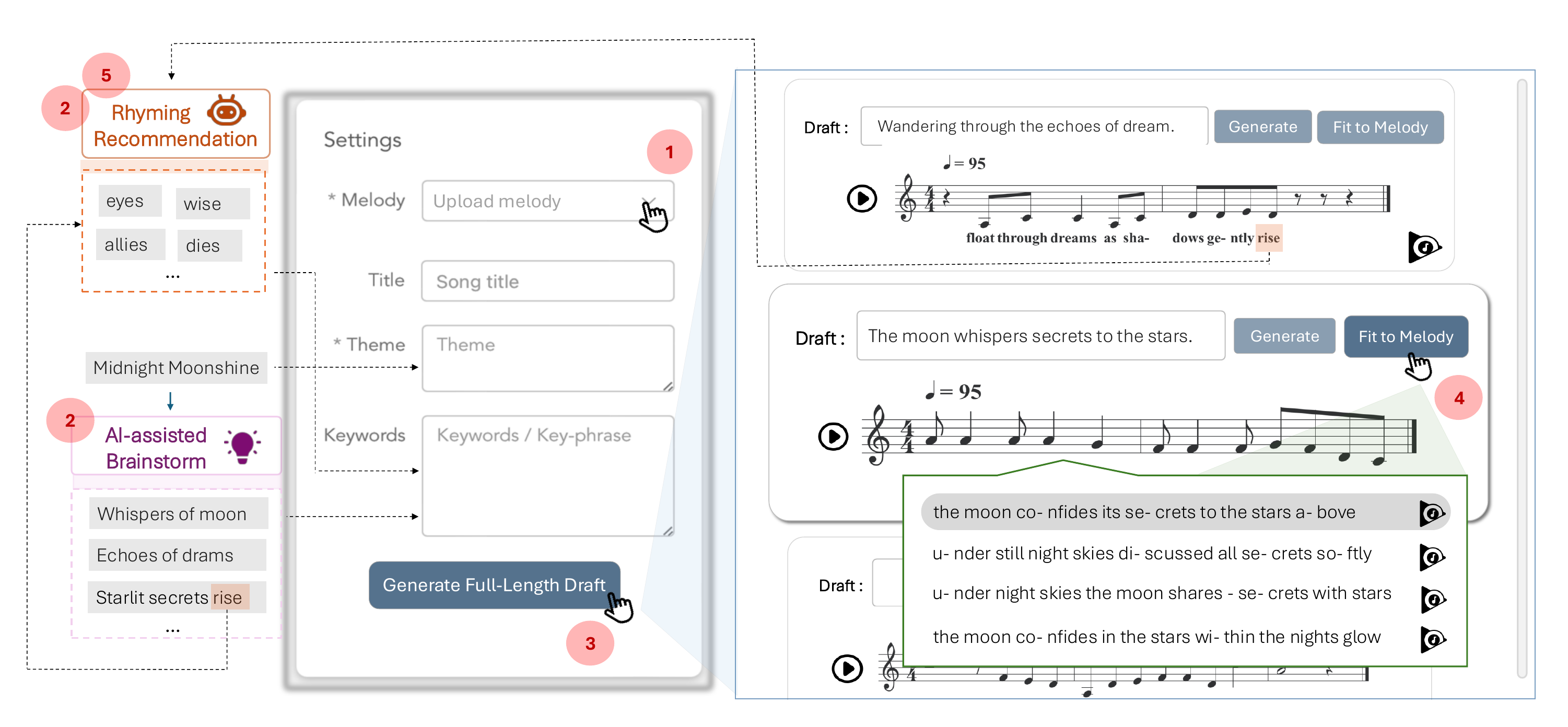}
  \caption{ 
  CoLyricist is an AI-assisted lyric writing tool designed to align with lyricists' typical workflows. The process involves the following steps: Users upload a melody and specify the song's theme (1).They then brainstorm ideas based on the theme, using the AI-Assisted Brainstorm function to generate relevant phrases and the Rhyming Recommendation function to get relevant rhyme suggestions. They can store these idea in a keyword/key-phrase memo box (2). Next, users draft lyrics, either manually or by using draft lyrics generation functions (3), which the output is influenced by both the theme and stored keywords. Users can use the Auto Melody-Fitting feature to adjust the draft lyrics to fit the melody line by line (4), while preserving their original meaning. Users can also use rhyming recommendations to further polish their lyrics (5). The tool supports an iterative process, enabling continuous refinement with AI assistance.
  }
  \Description{
  Screenshot of the CoLyricist user interface illustrating a workflow-aligned lyric writing process. The interface is divided into three main panels. On the left is the Settings Panel, where users upload a melody and specify the song theme and keywords, corresponding to the theme-setting stage (1). Below it is the Ideation Panel, which displays AI-assisted brainstorming phrases and thematically relevant rhyming suggestions that users can save as key phrases, corresponding to the ideation stage (2). On the right is the Editing Panel, where users draft lyrics and view musical notation aligned with the uploaded melody. Users can generate draft lyrics either for the full song or line by line, influenced by the selected theme and saved keywords, representing the drafting stage (3). An auto melody-fitting function allows users to adjust draft lyrics to fit the melody line by line, producing multiple candidate lyrics aligned with musical notes (4). Rhyming recommendations are also shown alongside the drafts to support further refinement during revision (5). Numbered markers, arrows, and dashed lines indicate an iterative workflow in which users repeatedly move between ideation, drafting, and melody fitting with AI assistance.
 }
 \label{fig:tool_design}
\end{teaserfigure}

\maketitle

\section{Introduction}

With the rapid advancements in large language models (LLMs), there has been a surge in HCI research focused on applying LLMs to assist in human writing activities \cite{Lee2022-mr, Jakesch2023-ih, Kim2024-ou, Dang2022-te}. These studies have confirmed the effectiveness of LLMs in enhancing the creative process. However, in the domain of music, while research on music 
~\cite{Zhu2023-qe, Frid2020-fz, Kim2025-xf} and lyric ~\cite{Zhao2024-ic, Tian2023-di, Zhang2024-fz, Chae2025-gh} generation models are emerging, there is a noticeable lack of studies examining how to leverage the recent development of generative models to assist song writing during the music creation process. Specifically, research on the interaction design for lyric writing support tools is sparse, despite the crucial role that lyric writing plays in song composition.

Lyric writing differs significantly from other writing tasks, such as essay or story writing, due to the necessity of aligning the lyrics with a melody \cite{Tian2023-di}. This requirement introduces unique challenges that cannot be directly addressed by applying insights from LLM-assisted writing in other domains. Therefore, there is a need for specialized research into the design of tools that can effectively support the lyric writing process.

Existing research on interactive lyric writing support tools \cite{Watanabe2017-as, Rongsheng2020Youling, Abe2012-zt} is limited and often lacks a deep understanding of the typical workflows and challenges faced by lyricists. In particular, although lyric writing is typically an act of personal emotional expression, these tools usually allow users to specify only coarse-grained themes and lack the ability to suggest lyrics that reflect the user's detailed messages while maintaining alignment with the melody. Moreover, many tools skip preparatory processes such as ideation \cite{lyricstudio, AISongLyricGenerator}, while others focus exclusively on supporting ideation \cite{masterwriter, lazyjot}, leading to environments that fail to offer a comprehensive and seamless lyric writing experience.


We introduce CoLyricist, an AI-driven lyric writing assistant for lyricists that addresses these gaps by providing comprehensive, workflow-aligned support across all major stages of lyric writing. By offering seamless AI-driven support tailored to multiple stages of the lyric writing process, CoLyricist seeks to enhance the cohesiveness and effectiveness of the lyric writing experience. Notably, it features a novel auto Melody Fitting interaction, where users draft free-form lyrics and transform them to match a melody, enabling them to convey their detailed message in the lyrics while preserving alignment with the musical structure.

To build CoLyricist, we first conducted a formative study involving interviews with 10 experienced hobbyist lyricists to understand their workflows and the challenges they face in lyric writing. The interviews revealed a diverse range of lyric writing workflows, yet a common structure emerged: (1) Theme setting, where the overall theme of the song is established; (2) Ideation, involving the collection of keywords and phrases related to the theme; (3) Lyrics Drafting, where the theme is refined and specific messages are decided for each section or melody phrase; and (4) Melody Fitting, the process of adapting the intended message to fit the melody.

Notably, most existing tools skip steps 2, 3 and 4, instead directly generating lyrics based on the theme and some musical constraint \cite{Watanabe2018-rt, Yamashita2022-pw, Abe2012-zt}. This approach often makes it difficult for creators to accurately reflect their intentions in the generated lyrics. The study also highlighted that \textbf{one of the most challenging tasks is Melody Fitting}, where the lyric must be aligned with the melody while maintaining meaning, syllable count, phonemes, and rhymes—making it a highly complex task.

\projectname{} is built on these insights, designed to provide seamless lyrics writing experience for lyricists.
We implemented each feature using existing large language models (LLMs) \cite{Zhao2024-ic, openai_gpt3_5}, as the scope of our research focuses on interaction design. Unlike traditional tools, \projectname{} is explicitly designed to align with the typical workflow of lyricists, supporting the crucial steps of Ideation, Lyrics Drafting, and Melody Fitting. We chose not to provide support for Theme setting, as the Formative Study indicated that participants generally did not express significant difficulties at this stage.

To explore how users interact with \projectname{} and to gain insights that guide the future design and development of a lyric writing support tool, we conducted an exploratory user study with 16 participants.
Half of the participants were experienced hobbyist writers with some previous experience in lyric writing, while the other half were novices with no prior lyric writing experience but possessed basic musical knowledge and an interest in lyric writing. Including novices allowed us to examine whether a tool designed based on workflows extracted from experienced lyricists could also be effectively applied to novice users. Participants were asked to use \projectname{} to write lyrics for a pre-selected melody.

The results showed that most participants found the tool helpful in streamlining the lyric writing process and were satisfied with the final lyrics, which accurately reflected their intended messages. The Melody Fitting feature was particularly beneficial for novices, while experienced lyricists found the Ideation support most useful. Additionally, we invited professional experts to evaluate the quality of created lyric. The evaluation result revealed that, given time constraints, novice writers using the \projectname{} produced lyrics of comparable quality to experienced writers, particularly in melody alignment and overall quality, suggesting that \projectname{} successfully narrows the gap between novice and experienced lyricists.

In summary, this research makes three key contributions:
\begin{enumerate}[noitemsep, topsep=0pt]
\item Formative Study: We conducted interviews with 10 experienced lyricists to clarify their workflows, identify challenges in lyric writing, and highlight limitations in existing AI services.
\item Design and Implementation of \projectname{}: We propose and implement a novel lyric writing support system, CoLyricist. Compared to existing systems, CoLyricist is characterized by the following two features: \textbf{i) a design that provides comprehensive AI support aligned with the typical songwriting process}, and \textbf{ii) a distinctive Melody Fitting interaction} that allows users to iteratively transform their drafted lyrics to fit a given melody.
\item User Study: We validated the effectiveness of \projectname{} design through a User Study involving 16 experienced and novice users, offering insights into the tool's impact and potential directions for future development.
\end{enumerate}



\section{Related Work}










Our research focuses on augmenting lyric writing with AI, a field that intersects with creative content generation and human-computer collaboration. This work builds upon prior research in three key areas: existing lyric writing practices, interactive tools for supporting lyric writing, and AI lyric generation models.

\subsection{Lyrics Writings Practice}\label{sec:related_work_lyrics_writing}
Lyric writing methods vary widely, reflecting the diverse approaches used by songwriters. A key distinction lies between the Melody-First and Lyrics-First approaches \cite{pamela-aa, jason-aa, Negus2015-bd, Reinhert2019-ue}.

In the Melody-First approach, the melody is composed first, with lyrics crafted to fit its structure. Oscar Hammerstein described this as “the American songwriter’s habit of writing the music first and the words later” \cite{Oscar-aa}. Similarly, Brian Eno improvises sounds while listening to backing tracks, later transforming them into meaningful lyrics \cite{Sheppard2009-aa}. Conversely, the Lyrics-First approach begins with independently written lyrics. For example, Hal David and Burt Bacharach sometimes prioritized lyrics when designing melodies \cite{Zollo-aa}, and Townes Van Zandt emphasized the importance of standalone lyric quality, stating, “words had to work on paper, without guitar” \cite{Negus2015-bd}. While there is no clear consensus, a user study on grief songwriting by Thomas et al. found that most participants favored the Melody-First approach \cite{Dalton2006-fi}. Both approaches require aligning lyrics with the melody, a process often involving trial and error to refine syllable counts and prosody \cite{Tommy-aa, jason-aa}.

Regarding the process of lyrics writing, theme setting is widely regarded as the initial step in lyric writing \cite{pat-aa, andrea-aa, pamela-bb, Takamitsu-aa, Tatsuji-aa}. Themes often emerge from personal experiences or emotions and frequently inform the song’s title. Techniques like Noun-Collision, which pairs random nouns to spark creativity, can help when themes are not immediately apparent \cite{Reinhert2019-ue}.

After determining the theme, lyricists differ significantly in their preparatory processes. Some emphasize ideation, brainstorming related words and phrases. Pat Pattison suggests compiling lists of thematic words and their rhymes using tools like thesauruses and dictionaries \cite{pat-aa}. Others focus on outlining, as Jason Blume and Stolpe recommend, by planning the content of each verse and chorus or establishing a plot progression for narrative consistency \cite{jason-aa, andrea-aa}. Shimazaki and Izutsu advocate for creating detailed character scenarios and emotional contexts to ensure thematic coherence, including details such as the protagonist’s background and emotional state \cite{Takamitsu-aa, Himi-aa}. Conversely, some lyricists bypass these steps altogether, favoring a more intuitive, inspiration-driven approach. Miranda Cooper, for example, mentioned that she places greater emphasis on intuition in her process  \cite{Kardzha2024-vx}.

In summary, lyric writing methods are highly diverse, with notable differences in preparatory processes such as ideation, outlining, and character development. Yet, it remains unclear which approaches are most commonly adopted or where lyricists encounter difficulties. Furthermore, little is known about the cognitive processes that bridge these preparatory activities and the actual act of writing, leaving a gap in understanding how ideas are transformed into lyrics.




\subsection{Interactive Lyrics Writing Assistants}

AI-powered writing assistants have greatly advanced creative writing in domains such as storytelling and poetry \cite{Weber2024-ek, Yuan2022-ds, Wu2022-nz, Goncalo_Oliveira2017-lb, Dhillon2024-sx, Reza2024-os}. \citet{Lee2024-hb} proposed a comprehensive design space for intelligent and interactive writing assistants, providing a valuable foundation for exploring system design. However, this framework remains abstract and general-purpose, offering limited guidance for specialized domains like lyric writing, which requires domain-specific interaction and structural considerations.

For instance, a number of lyric writing support tools have been proposed where AI generates and suggests lyrics, yet even within this category, the forms of interaction vary widely. These inputs can be broadly categorized into two types: those related to semantics and those related to format. \citet{Abe2012-zt} developed a system that proposes one line of lyrics at a time, allowing users to control only format-related aspects such as syllable count, rhyme, and word accent. In contrast, \citet{Watanabe2017-as} proposed a system that allows users to control not only syllables but also the semantic content by selecting coarse-grained themes at the word level from a set of presets. Furthermore, Youling, a system proposed by \citet{Rongsheng2020Youling}, enables users to specify emotional tone, thematic keywords (semantics), as well as constraints like acrostics, rhyming, and words-per-line (format). Meanwhile, commercial web services allow users to provide prose-style descriptions of the overall theme or mood of a song \cite{lyricstudio, AISongLyricGenerator}.

Lyric writing is often a deeply personal act of emotional expression. Nevertheless, existing systems tend to offer only coarse-grained control over semantic aspects—for example, by allowing users to specify the overall theme of the song in prose. As for format-related control, while some systems allow users to constrain syllable count, however, it is known that singability is not determined solely by matching syllable count but is also affected by the flow of the melody and the natural intonation of words \cite{Nichols2009-tp}. To address this gap, \textbf{CoLyricist introduces a novel Melody Fitting interaction}, designed based on insights from formative studies. This feature enables users to revise their line level message (draft lyric) to match a given melody while preserving their intended message, supporting the nuanced interplay between linguistic content and musical structure.

In addition to suggesting lyrics, writing assistants can also support preparatory processes. For example, \citet{Settles2010-wm} introduced Titular, a tool that generates novel song titles to serve as creative starting points, and LyriCloud, which presents a word cloud of semantically related terms to inspire users. \citet{Yamashita2022-pw} proposed the Lyric Association Map, a mind map tuned for multi-user collaborative lyric ideation. Online services have also emerged that allow efficient searching of rhyming words and related terms for lyric writing \cite{masterwriter, lazyjot}.

However, these systems remain fragmented, focusing either on preparatory ideation or lyrics drafting, without offering a seamless integration between the two. 
Meanwhile, in music composition, researchers have increasingly focused on analyzing the creative process itself to understand how human and AI contributions evolve across stages  \cite{Morris2025-ki, Fu2025-ui}.
For example, empirically examined how the introduction of AI reshapes the songwriting process and identified a new creative stage termed Collaging, where creators combine and refine multiple AI-generated ideas \cite{Fu2025-ui}. 
They further proposed a framework for understanding the evolving relationship between human agency and AI assistance in co-creative music production. 

As discussed in the previous subsection, lyric writing similarly involves a highly complex and iterative process encompassing multiple cognitive phases. 
Therefore, this work aims to deepen our understanding of the lyric writing process itself and to design interaction techniques that \textbf{seamlessly support diverse phases of lyric writing} within an integrated, workflow-aligned environment.

\subsection{Lyrics Generative Models}
Parallel to the development of interactive writing assistants, significant advancements have been made in lyrics generative models. Some earlier works primarily focused on generating lyrics without incorporating structural or melodic constraints such as syllable count or melody information \cite{Malmi2016-dl, Potash2015Ghostwriter, nikolov2020rapformer}. For example, \citet{Potash2015Ghostwriter} developed Ghostwriter, a system for generating rap lyrics using Long Short-Term Memory (LSTM) networks. Similarly, \citet{nikolov2020rapformer} proposed Rapformer, a conditional lyric generation model based on denoising autoencoders. Both models focus on textual aspects such as rhyming and word patterns, without incorporating melodic structure into the generation process.

Meanwhile, the intricate relationship between lyrics and melody is a fundamental aspect of songwriting, as thoroughly explored by \citet{Nichols2009-tp}. Their study revealed that lyrics and melodies are deeply interconnected\textemdash rhythmic and phonetic characteristics of lyrics often align with melodic contours. This highlights the importance of incorporating melodic context in lyric generation.

In response to this insight, recent research has increasingly focused on melody-aware lyric generation \cite{Lee2019icomposer, Qian2022Training, Ram2021-hl, Vechtomova2023-pe, Vechtomova2021-ld, Ma2021-ga, Oliveira2015-yc, Sheng2021-nq}. For instance, \citet{Watanabe2018-rt} introduced a pioneering melody-conditioned lyrics language model that demonstrated how lyric generation can be guided by melodic structures. Building on this foundation, \citet{Ding2024-gn} proposed SongComposer, a large language model capable of generating both lyrics and melody simultaneously.

Exploring a different direction, \citet{Tian2023-di} presented Lyra, an unsupervised model for melody-to-lyrics generation that does not require paired training data. Building upon Lyra, \citet{Zhao2024-ic} proposed REFFLY, a melody-constrained lyrics editing model that transforms user-written prose into lyrics aligned with a given melodic structure, while preserving the original meaning as much as possible. Recognizing REFFLY ability to directly incorporate melodic constraints and provide fine-grained, phrase-level lyric control, we adopt REFFLY as a component within CoLyricist to support our system melody-fitting functionality.

\begin{figure}
    \centering
    \includegraphics[height=6cm]{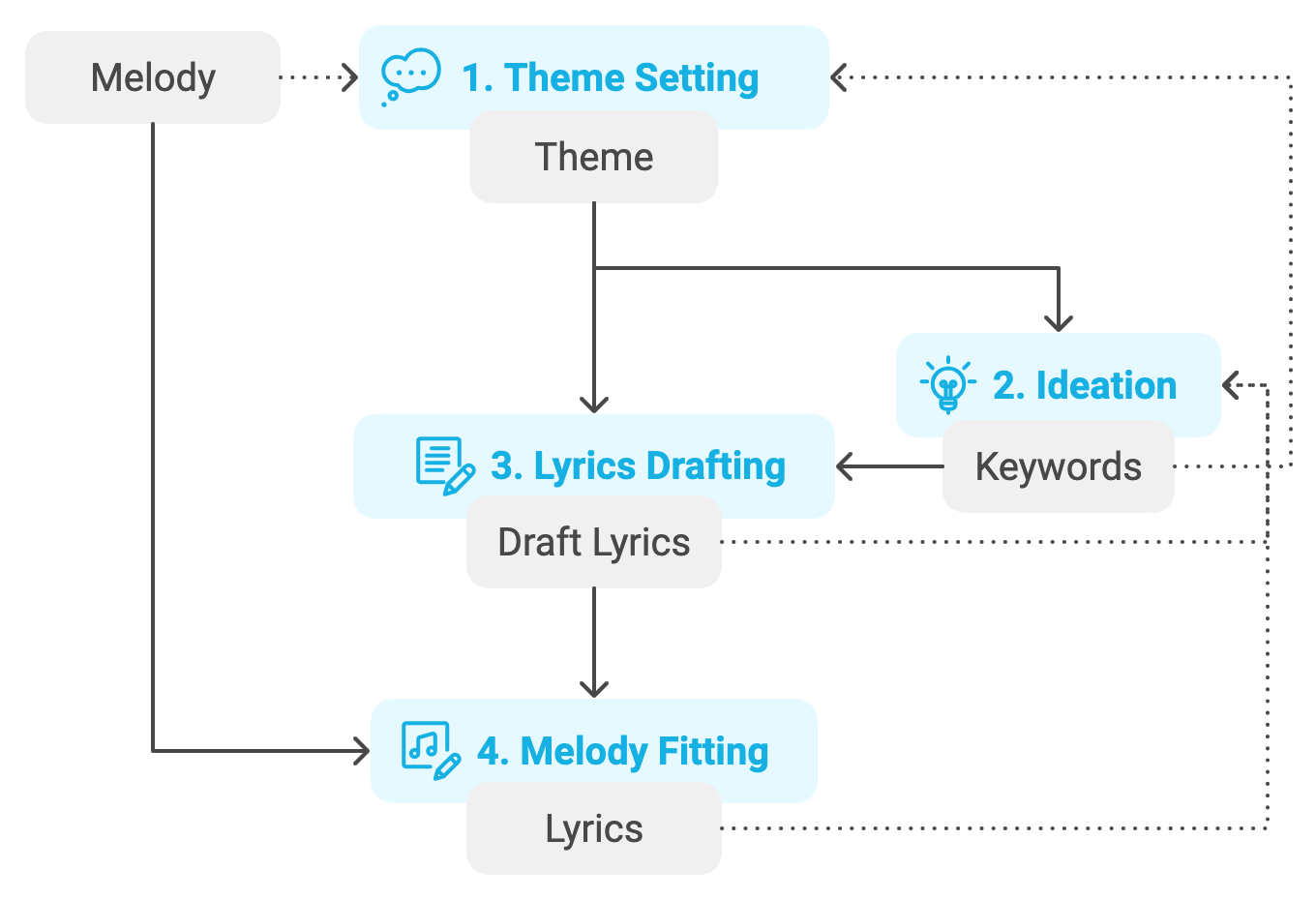}
    \caption{High-level typical lyric writing workflow, comprising four primary stages: Theme Setting, Ideation, Lyrics Drafting, and Melody Fitting. Gray boxes represent stage outputs, while blue boxes denote actions.}
    \Description{Diagram illustrating a high-level lyric writing workflow composed of four stages arranged vertically. At the top is “Theme Setting,” which receives input from the melody and produces a theme. To the right is “Ideation,” which takes the theme as input and generates keywords. Below is “Lyrics Drafting,” where draft lyrics are created based on the theme and keywords. At the bottom is “Melody Fitting,” where draft lyrics are adjusted to fit the melody, producing final lyrics. Blue boxes represent user actions or workflow stages, while gray boxes represent intermediate or final outputs. Solid arrows indicate the primary progression from theme setting to ideation, drafting, and melody fitting. Dotted arrows indicate iterative feedback loops, showing that users can revisit earlier stages such as ideation or drafting after melody fitting.
}

    \label{fig:workflow}
\end{figure}

\section{Formative Study}
As discussed in Section \ref{sec:related_work_lyrics_writing}, existing literature presents a wide range of approaches to lyric writing. In particular, the steps taken after setting a theme are highly diverse. This diversity makes it unclear what workflows and challenges lyricists typically face in practice. To gain deeper insights into the typical workflow and challenges they encounter, we conducted a formative study involving semi-structured interviews with 10 experienced hobbyist lyricists. In this section, we present the main findings and challenges identified through the interviews, focusing on those that significantly influenced the design of our system. These are labeled as \textbf{C1–C5} for challenges and \textbf{F1–F4} for other findings.


\subsection{Interviewees and Analysis Process {Interviewees \& Process}}
We conducted interviews with 10 individuals who have experience in music creation as a hobby (I1-I0). All participants were aged between 20 and 30, with 6 males and 4 females. 6 of the participants were university students, while the remaining 4 were employed in industries such as electronics manufacturing and software development. 
In terms of lyrics writing experience, 4 participants had less than 3 years, 5 had up to 5 years, and the remaining 1 had over 5 years. Regarding the number of lyrics they have written, 4 participants had written fewer than 10 songs, 4 had written between 10 and 20 songs, and 2 had written over 20 songs. Additionally, 7 participants had experience posting their self-written songs on social media or music streaming platforms, and 5 had written lyrics to perform with their amateur bands.
The interviews were conducted via online video calls, with each session lasting approximately 45 minutes.


The interviews were structured around four main sections. First, we asked participants about their background and experiences in lyric writing and composition. Second, we explored their typical songwriting workflows, with a particular focus on the process of writing lyrics. Third, participants were asked to reflect on the aspects of lyric writing that they found most challenging or time-consuming. Finally, we discussed participants’ use of AI tools in their creative practice, including the services they had used and their perspectives on the strengths and limitations of these tools. A list of example interview questions is provided in the Appendix \ref{app:interview_protocol}.

 The interview recordings were fully transcribed and analyzed using an inductive qualitative analysis approach. The primary goal of this analysis was to identify recurrent patterns in participants’ descriptions of lyric-writing workflows and challenges that could inform system design. Specifically, the first author carefully read all interview transcripts and highlighted segments related to how participants plan, develop, and refine lyrics. These segments were iteratively examined and grouped to capture common patterns across participants’ practices. Although the primary analysis was conducted by the first author, the emerging categorizations and resulting workflow model were regularly discussed with the co-authors to reduce interpretation bias and ensure consistency with participants’ accounts. In Section \ref{sec:workflow}, we present the identified workflow in detail.

\subsection{Lyric Writing Workflow} \label{sec:workflow}
This section outlines the typical high-level workflow of lyric writing, as identified through our interviews. While there was significant diversity in the details of each participant's process, several common steps emerged at a high level. We have summarized this typical high-level workflow in Figure \ref{fig:workflow}. It is important to note that this figure represents the most common workflow, but even at this high level, there is variability among individuals. For instance, some creators skip certain steps, while others follow the process implicitly without explicitly recognizing each step. The remainder of this subsection will describe the typical workflow and its challenges based on Figure \ref{fig:workflow}.

\paragraph{\textbf{Step1. Setting the Theme:}} This step involves establishing the overall theme of the song. The theme provides coherence throughout the song but is not necessarily the message the song aims to convey—it could also be a situation or a scene. For example, some participants indicated that their songs do not carry a strong message but instead depict scenes or emotions that inspired them.
Even when the theme is not explicitly verbalized, it exists in the creator's mind. When verbalized, the theme is usually expressed in one or two sentences. 

There are three main methods for setting the theme. The first is through personal inspiration, where they note down strong feelings or thoughts during everyday life and later selects a theme from these notes. I9 noted \textit{"I usually notice something small and note it down in everyday life, like a scene I find interesting, and that feeling becomes the theme of the song."} The second method is melody-inspired, where a theme is set based on emotions or imagery evoked by a pre-composed melody. I7 mentioned \textit{"Often the theme is not something I clearly decide in advance. I imagine a scene from the melody, and that image itself becomes the theme"}. This method is taken only in melody-first writing style. Lastly, in the case of commissioned work, the theme is often provided by the client, such as when a song is requested for a specific event.

It was observed that \textit{most participants did not struggle with setting the theme} \textbf{ (F1)}, as it often serves as the motivation for starting the songwriting process. I8 noted \textit{"The theme itself is what motivates me to start writing"}. This may be especially true for hobbyist lyricists, who, without strict deadlines, can begin writing whenever inspiration strikes—a flexibility that may not be available to professional writers.


\paragraph{\textbf{Step 2. Ideation:}} Before beginning to write lyrics, participants gathered keywords and phrases related to the theme. Many referred to thesauruses or conducted online searches to collect related words and phrases. The collected material primarily fell into two categories.
First, participants sought out "unique word combinations" that aligned with the theme. While finding related individual words using a thesaurus was relatively straightforward, \textit{finding unique and meaningful word combinations or phrases that were relevant to the theme proved to be a challenge for them} \textbf{(C1)}. 
I10 mentioned \textit{"When there is a theme, even coming up with words related to it is already difficult. And finding interesting combinations of those words is even harder"}. 
Second, once participants identified key phrases or keywords they wanted to use, they searched for rhyming words to incorporate into the lyrics. While basic rhyme dictionaries were useful for finding rhyming words, \textit{finding rhymes that were also relevant to the theme posed them additional difficulty} \textbf{(C2)}.
I3 noted \textit{"When looking at rhyme books, you often find that only a small set of commonly used words are actually practical, and choosing the right ones from this limited pool to fit my theme is tough"}. 
This step occasionally led to further refinement and specification of the theme. However, it is worth noting that not all participants explicitly followed this step, as some preferred to skip ideation entirely and proceed directly to writing lyrics.

\paragraph{\textbf{Step3. Drafting lyrics from the Theme}:} This step involves breaking down the theme into specific messages for each melody phrase, using the collected keywords and phrases. These initial drafts, referred to as ``Draft Lyrics'' in this paper, may directly incorporate phrases generated during ideation. 

There are \textit{two main approaches to this lyrics drafting process} \textbf{(F2)}. The first is the full-song perspective, where some participants develop a complete draft of the lyrics before moving on to the next step, Melody Fitting. Participants who prioritize narrative structure tend to follow this approach, outlining the message for each section and then further breaking it down for each melody phrase. I3 mentioned \textit{"I usually have a general outline and a main idea, but the song is made up of multiple scenes or segments that come to mind, almost like a slideshow"}. The second approach is phrase-by-phrase, where others combine this step with Melody Fitting, creating the draft and adjusting it to the melody one phrase at a time.  I6 noted \textit{"I often start by writing phrases that fit the melody in an ad hoc, improvisational way"}. These participants may not explicitly write out the Draft Lyrics but still follow this process internally. Notably, only two of the interview participants followed the former approach, where they first developed the song's overall structure before addressing individual phrases. In this step, some participants noted that \textit{while some ideas, such as the hook phrase, often came quickly once the theme was established, generating ideas to fill the remaining sections of the song proved to be challenging} \textbf{(C3)}. Specifically, I4 noted \textit{"After creating the melody, I often identify one or two lines that evoke a strong feeling. [...] I struggle to fill in the other parts of the lyrics"}.

\paragraph{\textbf{Step4. Melody Fitting:}} The draft lyrics were then adjusted to fit the melody while maintaining their meaning. Melody constraints include factors such as syllable count, prosody, and rhyming. Participants often searched for rhyming words at this stage again. Most participants confirmed the compatibility of the lyrics with the melody by singing them, while some used commercial vocal synthesizers to check this. Those who used vocal synthesizers were usually not the lead singers in their bands or did not typically sing themselves. \textit{Participants who were less accustomed to singing noted that even when they sang the lyrics themselves, objectively evaluating the singability was particularly challenging} \textbf{(C4)}. I6, who does not serve as the vocalist in the band, stated: \textit{“When I want to check how the melody and lyrics match, I map the syllables in a music production software and try singing them. Even then, there are times when I can’t properly evaluate it, so I occasionally use a singing voice synthesis tool to check”}.

Eight out of ten participants mentioned in the interviews that the Melody Fitting process was considered the most challenging aspect of lyric writing. \textit{The complexity of maintaining the message while considering syllable count, prosody, and rhyming made this task particularly difficult} \textbf{(C5)}. P3 said \textit{"The challenge is turning a clear idea into lyrics that remain accessible and poetic, while also making sure they can be sung well and fit constraints such as rhyme and rhythm—it feels like dancing with shackles"}. Participants indicated that their current approach involves either waiting for inspiration to strike or searching for references in songs or literature to guide them through this process.

\vspace{10pt}

Regardless of whether participants followed a Melody-First or Lyrics-First approach, the high-level workflow was largely consistent (Seven of the participants mentioned they often take the Melody-First style). However, those who followed a Lyrics-First approach were more likely to write a refined Draft Lyrics in this step and then adjust it in the Melody Fitting step.

Interestingly, \textit{Participants do not always start writing from the beginning of the lyrics} \textbf{(F3)}. Many participants preferred to start with the hook or chorus, the most memorable part of the song. While the hook was often determined quickly, writing the rest of the lyrics was often more difficult, as they needed to maintain coherence with the existing lyrics while also meeting melody constraints.

To summarize, while Section \ref{sec:related_work_lyrics_writing} discusses that the processes following theme setting may include ideation, outlining, and character scenario development, \textbf{our interviews revealed that the typical processes after setting a theme are primarily "Ideation". Moreover, we found that transforming the generated ideas into actual lyrics typically involved, either explicitly or implicitly, a two-step process consisting of Lyrics Drafting and Melody Fitting. Melody fitting was explained as the most challenging aspect.} 

Regarding character scenario development, some participants noted that character scenarios emerge by the time their lyrics were completed, but none reported creating detailed story settings in advance of writing the lyrics. The Lyrics Drafting step, for some participants, involved creating an outline for each section of the song before deciding on the specific message for each melody phrase. In this context, outlining can be seen as a subset of the Lyrics Drafting step. 

\subsection{Use of AI Tools}
When asked about their experience with AI tools for lyric writing, 7 participants had tried conversational AI services like ChatGPT\footnote{\url{https://chat.openai.com}}, and 4 had experimented with Suno AI, an AI-based automatic music generation service that also includes lyric generation. None of the participants had used specialized lyric-writing support tools.

\subsubsection{Controllability Issues with Existing AI Tools}
Many participants expressed difficulty using existing AI services in their lyric writing due to limited controllability. For instance, Suno AI\footnote{\url{https://suno.ai}} generates complete lyrics based on a broad theme but does not allow for finer control. Regarding this point, I2 said, \textit{“If we disregard technical limitations, I would prefer a tool that helps me refine each line of lyrics”}. While ChatGPT can theoretically accommodate more detailed instructions, this requires highly crafted prompts, making it unsuitable for efficient lyric writing. Moreover, neither service supports generating lyrics with melody constraints. As a result, participants rarely adopt generative AI services in their workflow.

\subsubsection{Need for Workflow-Aligned Support}
During the interviews, we also discussed the use of Suno AI for music composition. Suno AI allows users to input a general style and then generates a song as an audio file. While this paper does not delve into the music composition process, it is worth noting that Suno AI's UX bypasses all the initial and intermediate steps of composition, outputting only the final result. Consequently, while users found it might be useful for generating commercial music or background music, none could integrate it into their creative process. This finding \textit{highlights the importance of providing support aligned with the creator's workflow} \textbf{(F4)}.

\subsection{Design Decision}
Based on the findings from the formative study, we established the following design decision:

\begin{enumerate}[label=\textbf{D\arabic*:}, leftmargin=*, align=left]
    \item \textbf{Provide AI Support Aligned with the lyricist's Workflow} 
    
    The tool should support the typical workflow of lyricist, as lyric writing tools that do not align with natural workflows are difficult to integrate into the creative process \textbf{(F4)}. Therefore, CoLyricist allows creators to work through each stage of the lyric writing process while providing AI-driven assistance at key steps. This approach aims to enhance the productivity of lyricists by offering support without disrupting their workflow.
    
    \item \textbf{Facilitate Ideation} 
    
    To facilitate the ideation process, the tool offers two key functionalities. First, it enables users to brainstorm not only single words but also unique phrases or combinations of words that are closely related to the theme \textbf{(C1)}. Second, the tool enhances the search for rhyming words by prioritizing those that align with the specified theme, rather than providing random rhyming suggestions like traditional rhyming dictionaries \textbf{(C2)}. By incorporating these tasks directly into the tool, it streamlines the workflow and minimizes disruptions, improving the overall efficiency of the lyric-writing process.
    
    \item \textbf{Assist with Drafting Lyrics} from the Theme 
    
    The tool should help users break down the theme and generate draft lyrics. This process should incorporate the keywords gathered during the Ideation phase. The tool must provide flexible support to accommodate different working styles \textbf{(F2)}, offering both the option to generate lyrics for the entire song at once or one melody phrase at a time. The phrase-by-phrase generation feature should be particularly valuable for users facing challenges, such as filling in less-developed sections when some other parts are already established \textbf{(C3)}. It also supports users who wish to start writing from a specific part of the song \textbf{(F3)} or revisit and revise previously written lyrics.
    

    \item \textbf{Support Melody Fitting} 
    
    The tool should assist users in refining their draft lyrics to align seamlessly with the melody while preserving the original intent. As Melody Fitting is considered the most challenging step in lyric writing \textbf{(C5)}, providing robust support at this stage is expected to significantly enhance productivity. To enable users objective evaluation of the suggested lyrics, the tool incorporates a vocal synthesis feature, allowing users to hear how the lyrics sound when sung to the melody \textbf{(C4)}. Additionally, the tool should allow users to flexibly adjust the proposed lyrics while maintaining adherence to melodic constraints, ensuring that creativity and precision are not compromised.

\end{enumerate}

\begin{figure*}
    \centering
    \includegraphics[width=\linewidth]{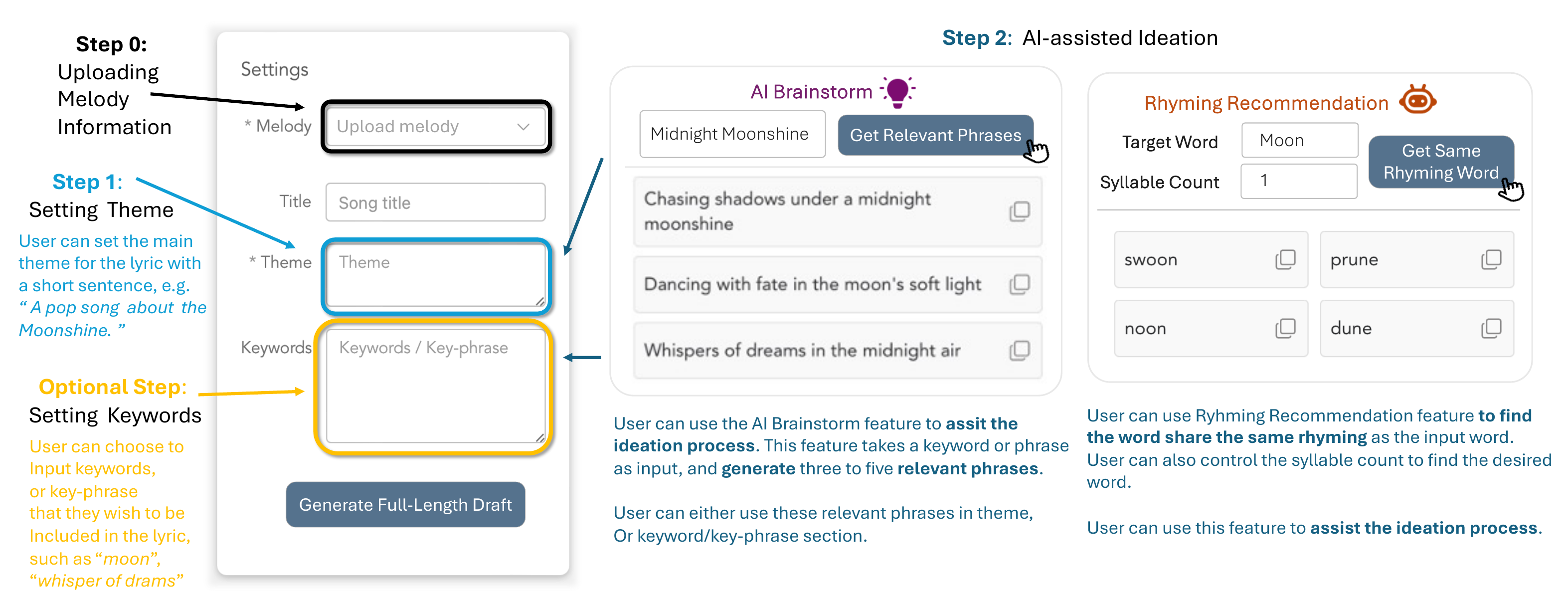}
    \caption{The first-half workflow: Step 0 (uploading melody information), Step 1 (setting the theme), and Step 2 (ideation).}
    \Description{Annotated screenshot of the CoLyricist interface illustrating the first half of the lyric writing workflow. On the left is the Settings panel, where Step 0 highlights the melody upload field used to input melody information. Below it, Step 1 highlights the theme input field, where users enter a short sentence describing the main theme of the song. An optional keyword or key-phrase input field is also shown, indicating that users may specify words they wish to incorporate into the lyrics. On the right is the ideation area corresponding to Step 2, which includes two AI-assisted features: the AI Brainstorm panel that generates several thematically relevant phrases from a given keyword or phrase, and the Rhyming Recommendation panel that allows users to input a target word and syllable count to obtain rhyming word suggestions. Arrows, colored outlines, and textual annotations visually connect each interface element to its corresponding workflow step.}
    \label{fig:first_half_workflow}
\end{figure*}

\section{Design and Implementation}
\projectname{} is a lyric-writing support system designed to maximize both the efficiency and creativity of lyricists by aligning with typical lyric-writing workflows (D1). The system comprises three main components: the Setting Panel for configuring the song's theme and conditions, the Ideation Panel for assisting in brainstorming, and the Editing Panel for drafting and refining lyrics. In the following subsections, we explain how \projectname{} works in detail. Appendix \ref{app:tool_overview} illustrates the complete view of UI design of \projectname{}.

\subsection{Step 0: Uploading Melody Information}
\paragraph{\textbf{Design:}}
In the Setting Panel, users can upload a melody in MusicXML or MIDI format (Black highlighted in Figure \ref{fig:first_half_workflow}). For those who prefer the lyrics-first style, this step can be skipped initially and the melody can be uploaded after Step 3. Once the melody is uploaded, the Editing Panel displays the musical score, segmented into individual melody phrases (Green highlighted content in Figure \ref{fig:rest_workflow}). Users can listen to each melody segment by clicking the play button located to the left of the score.

\paragraph{\textbf{Implementation:}}
To segment the melody into typical phrase lengths, which generally correspond to one line of lyrics, the system automatically analyzes the uploaded file using music theory rules. The segmentation algorithm relies on rhythmic cues, such as longer notes or rests, which often indicate the boundaries of musical phrases. \cite{Knosche2005-perception}.

\subsection{Step 1: Setting the Theme}
\paragraph{\textbf{Design:}}
The first step in the typical lyric-writing process is setting the theme for the song. Users can optionally input a title, which often encapsulates the overall theme of the song. They then write the theme in free-form text, typically 1 to 3 sentences long. This length is based on common practices observed among participants in formative study. For example, a theme might be something like "A pop song about the moon shine." Since setting the theme is generally not challenging for lyricists (F1), the system does not provide an AI support for this step.

\subsection{Step 2: Ideation (D2)}
\paragraph{\textbf{Design:}}
After setting the theme, the next typical step is ideation. This step aims to help users generate and refine ideas based on the theme while gathering phrases and words they want to incorporate into the lyrics. The "Keyword/Key phrase" box in the Setting Panel serves as a free-form editor where users can jot down any words or phrases they intend to use in the lyrics (\textit{Optional Step} in Figure\ref{fig:first_half_workflow}). These texts are used as an inputs to the subsequent step. To further assist in this ideation process, the Ideation Panel offers two key features (\textit{Step 2} in Figure\ref{fig:first_half_workflow}).

The first feature is called "Brainstorming." Once the theme is set, pressing the "Get Relevant Phrases" button generates related phrases. If users need ideas from a more specific domain, they can input a keyword to generate phrases associated with that domain. The second feature is the "Rhyming Recommendation." This tool allows users to input a target word, and it suggests rhyming words related to the theme, with an option to specify the number of syllables. These features, particularly the "Rhyming Recommendation," are also intended to be used when users refine the lyrics after Step 4.
If users find any of the suggested phrases or words appealing, they can copy them into the "Keyword/Key phrase" text box. 

\paragraph{\textbf{Implementation:}}
Both the Brainstorming and Rhyming Recommendation features are implemented using the ChatGPT API, leveraging few-shot prompting to generate relevant suggestions. The prompts are designed to incorporate the user-provided theme, ensuring that the generated phrases and rhyming words align with the intended context. Details of the prompt design and implementation are provided in Appendix \ref{app:prompts}.

\begin{figure*}
    \centering
    \includegraphics[width=1\linewidth]{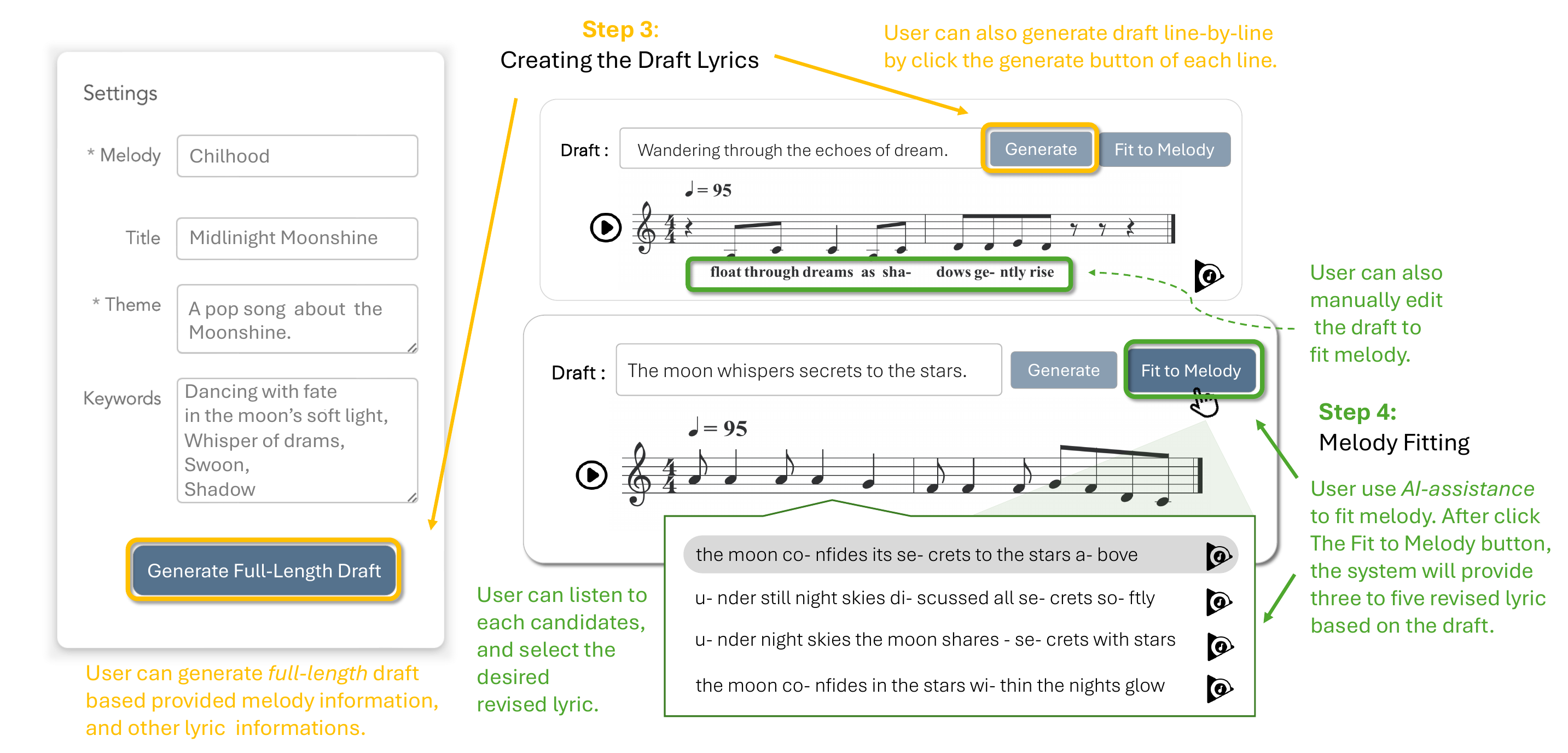}
    \caption{The rest half of the workflow: Step 3 (creating the draft lyric) and Step 4 (melody fitting) }
    \Description{Annotated screenshot of the CoLyricist interface illustrating the second half of the lyric writing workflow. Step 3 shows the draft lyrics creation stage, where users can generate a full-length draft based on the provided melody, theme, and keywords, or generate draft lyrics line by line using a generate button for each line. Each draft line is displayed alongside the corresponding melody staff, and users can listen to the melody playback. Step 4 shows the melody fitting stage, where users can manually edit draft lyrics to better match the melody or use the Fit to Melody function to receive three to five AI-generated lyric revisions aligned with the melody. The revised lyric candidates are presented in a list, each with audio playback controls, allowing users to listen to and select the desired version. Colored outlines, arrows, and annotations indicate the relationship between user actions, AI-assisted features, and the corresponding workflow steps.}
    \label{fig:rest_workflow}
\end{figure*}

\subsection{Step 3: Creating the Draft Lyrics \textbf{(D3)}}
\paragraph{\textbf{Design:}}
The next typical step in lyric writing is breaking down the theme into a Draft Lyrics. In \projectname{}, users can either write the Draft Lyrics themselves or opt for auto-generation. There are two auto-generation options (\textit{Step 3} in Figure \ref{fig:rest_workflow}).

The first option is generating the entire Draft Lyrics at once. If both the title and theme are set, pressing the "Generate Full-length Draft" button creates a Draft Lyrics for all melody phrases based on the title and theme. If neither is provided, a random theme is used. Additionally, any phrases or words entered in the "Keyword/Key phrase" text box are more likely to appear in the Draft Lyrics. This feature is intended for lyricists who prefer to conceptualize the entire song before moving forward.

The second option allows for generating a Draft Lyrics phrase by phrase. Pressing the "Generate" button next to the editing box for each melody phrase generates a draft for that specific phrase. Like the first option, the generated draft is influenced by the title, theme, and any keywords or key phrases provided. This feature is ideal for those who prefer to write lyrics one line at a time.

Users can also manually edit the generated Draft Lyrics. If the generated draft does not accurately reflect the message they want to convey in a particular melody phrase, they can modify it to better align with their intent.

\paragraph{\textbf{Implementation:}}
Both auto-generation options utilize the ChatGPT API with few-shot prompting techniques. Detailed prompts provided in the Appendix \ref{app:prompts}. The prompts are carefully crafted to incorporate the title, theme, and any keywords or key phrases entered by the user in earlier steps. The prompts include examples of lyrics, ensuring that the generated draft align with the typical length of a lyric line, which generally corresponds to the length of a typical melody phrase. Since the melody input in Step 0 is also segmented into typical phrase lengths, the generated draft is less likely to be unexpectedly too long or too short relative to the melody.

\subsection{Step 4: Melody Fitting \textbf{(D4)}}
\paragraph{\textbf{Design:}}
Once the Draft Lyrics is complete, the final step is fitting the lyrics to the melody. In this step, users can either write the lyrics themselves or use AI assistance. Below each note in the score, there are invisible text boxes. If they choose to write manually, they need only ensure that each box contains no more than one syllable, as it is uncommon in music for more than one syllable to be assigned to a single note \cite{Tian2023-di}. 

For AI assistance, users can press the "Fit to Melody" button next to each Draft Lyrics text box. This action generates revised lyrics that maintain the intent of the Draft Lyrics while fitting the melodic constraints. Notably, this feature works regardless of whether the Draft Lyrics is longer or shorter than the corresponding melody phrase. If the lyrics are too long, some content is trimmed; if too short, the content is automatically expanded. This means that users can input Draft Lyrics in a word-level style, even if they do not have a fully formed message in mind. The melody fitting feature targets melody-driven lyric writing styles and is intended to support a broad range of representative contemporary music genres, including jazz, country, blues, folk, pop, and comedy, consistent with the scope of the underlying melody-fitting model; styles primarily based on spoken rhythm (e.g., rap) are outside this focus.

Users can also listen to the generated lyrics by pressing the play button next to each suggestion to assess their singability. This means that they can evaluate the proposed lyrics based on both semantics and singability, selecting the one they prefer.

Once a set of lyrics is chosen, the syllables are displayed in the corresponding text boxes linked to the melody notes. Users can further edit these boxes to refine the lyrics, ensuring they better reflect their intended message while maintaining alignment with the melody. Users can also listen to the edited lyrics by pressing the play button placed under the score.

In practice, it is common for creators to have some lyric ideas already in mind. The tool is designed so that users can write their own lyrics where they have ideas and use AI assistance for the parts where they don't.

\paragraph{\textbf{Implementation:}}
To implement the Melody Fitting feature, we adopted the "Reffly" lyric generation model \cite{Zhao2024-ic}. Reffly is capable of revising text to fit a given melody while preserving the intended meaning. It also considers the context of previously written lyrics, ensuring consistency throughout the song. We re-implemented Reffly using GPT \cite{openai_gpt3_5} as a pre-trained model, instead of the original LLaMA2-13b model. Additionally, we designed mechanisms to enhance lyric revision capabilities. (More details are provided in Appendix \ref{app:implementation}.) 

The playback feature for lyrics is implemented using the "Sinsy" API, which synthesizes singing voices from melody and lyrics input in ABC Notation format \cite{Oura2010-fl}.

\section{User Study Settings}
This study evaluates CoLyricist as an integrated, workflow-aligned lyric-writing support system, with the goal of understanding how users engage with AI assistance across the entire lyric-writing process. Because the defined workflow serves as the foundation of the system’s design, a central focus of the study is whether CoLyricist can effectively support this workflow in practice. Accordingly, rather than conducting baseline comparisons (e.g., unaided writing or generic LLM-based tools), the study was designed as an exploratory investigation to surface design insights for future lyric-writing support systems.



The user study consists of two phases: Lyrics Writing Session and Lyrics Assessment Session. In Lyrics Writing Session, participants used CoLyricist to write lyrics for a given melody. This phase focused on observing whether users could follow the defined workflow smoothly and subjectively evaluating the usability of each function. These observations served to validate our design decisions and identify areas for improvement. In Lyrics Assessment Session, experienced lyricists evaluated the quality of the lyrics generated during the first session. This step aimed to confirm that lyrics created using CoLyricist maintain a certain level of quality and are suitable for actual use.

Participants included not only experienced lyricists but also novices. Since novices typically do not have well-established workflows, we sought to examine whether a workflow derived from experienced users—if sufficiently human-centered—could also be useful and intuitive for novice users. To verify this, we included novices in the study.

\subsection{Participants} We recruited participants through various Slack channels associated with multiple universities. An application form was shared within these channels, reaching groups ranging from dozens to hundreds of members. We specifically recruited two distinct groups for the study: 8 experienced lyricists who had written lyrics as a hobby \textbf{(P9-P16)}, and 8 novice participants with no prior lyric writing experience \textbf{(P1-P8)}, resulting in a total of 16 participants. The novice participants were selected based on a pre-study survey, which indicated they could read simple music scores and had an interest in lyric writing, despite never having written lyrics before. Among the experienced lyricists, 3 had experience writing over 10 songs, 2 had written between 5 and 10 songs, and 3 had written 5 songs or fewer. Participants were all students and graduate students from nine different institutions in the U.S., with academic backgrounds in fields such as Computer Science, Geometry, Biology, Fine Art, and Psychology (i.e., 8 females, 8 males, aged 18-30). All participants had some experience using ChatGPT.

The user study was conducted via Zoom, with participants accessing the CoLyricist tool in their web browsers. During the lyric writing sessions, participants shared their screens throughout, and microphones were muted to minimize distractions except when they had questions about the system.  The shared screens were recorded throughout the session. Participants were compensated with a \$30 gift card for completing the 60-90 minute session.

\subsection{Lyric Writing Session Procedure} Participants were tasked with writing lyrics to a pre-selected melody using CoLyricist. 
This user study evaluates the system within a melody-first workflow. While both melody-first and lyrics-first approaches are used in lyric writing, participants in our formative study frequently reported adopting a melody-first process and identified melody fitting as a key challenge, making this setting suitable for evaluating the system’s melody-fitting support.
The session followed the protocol described below:

\subsubsection{Introduction and System Walk-through (10 min)} Participants were introduced to the study and its background. They accessed CoLyricist and shared their screens while the facilitator guided them through the tool’s functionalities. The participants were asked to try each feature under the facilitator's instructions. Prior to the session, participants were also asked to watch a 6-minute tutorial video.

\subsubsection{Lyric Writing (~25 min × 2)} Participants completed two lyric-writing sessions, each using a different contemporary folk melody. One melody was in a minor key and the other in a major key, with different music characteristics, such as rhythmic density and rhythm complexity to confirm that CoLyricist could accommodate various musical styles\cite{Nick2024Exploring}. Participants were asked to firstly listened to the melody and then compose the lyrics. Each melody contained eight phrases (approximately 24 measures), covering the first verse and chorus, which are most popular song structures that exist in most of the popular songs \cite{Zhao2024-ic}. 

To motivate participants, they were given the following two lyric writing purposes for the two sessions: \begin{itemize} \item Purpose 1: Write a song to express gratitude to a loved one (family, partner, or close friend) for a special occasion (birthday, anniversary, Mother’s Day). \item Purpose 2: Write a song to perform at a university graduation party with friends, reflecting on your shared memories. \end{itemize}

Participants were free to determine their specific theme within these purposes and were instructed to imagine they would perform the song themselves. There were 2 melodies (major and minor) and 2 purposes, leading to 8 possible combinations. The order of the melodies and purposes was counter-balanced across participants to ensure each combination was covered evenly.

\subsubsection{Questionnaire and Interview (~20 min)} After the lyric writing sessions, participants completed a questionnaire in which they rated the usability of each feature of the tool, as well as their overall experience, on a five-point scale across several dimensions. A semi-structured interview followed, informed by their questionnaire responses and observed behaviors during the session.

\subsection{Lyrics Assessment Session Procedure} 
\label{sec:assessment-setting}
The Lyrics Assessment Session evaluated the quality of the lyrics created during the Lyrics Writing Session. Two external experts, who had not participated in the Lyrics Writing Session, were recruited as evaluators. Both evaluators are professional composers with extensive experience composing music across a variety of contemporary pop genres. Selected lyrics generated by users are provided in Appendix~\ref{app:lyrics}.



The session was conducted over Zoom. The facilitator shared a musical score aligned with the lyrics written during the Lyrics Writing Session and played a synthesized vocal performance of the song. The evaluators were asked to assess each song based on the following four criteria, each rated on a five-point scale (1. Very Poor, 2. Poor, 3. Neutral, 4. Good, 5. Very Good): 
\begin{itemize} 
    \item \textbf{Melody Alignment}: How well the lyrics fit with the melody in terms of rhythm, phrasing, and emotional tone. 
    \item \textbf{Creativity of the Phrases}: Originality and creativity of the lyrics. Are the lyrics fresh, engaging, and imaginative? 
    \item \textbf{Lyrical Coherence}: How well the lyrics flow together as a whole. Are the ideas and themes consistent and easy to follow? 
    \item \textbf{Overall Quality}: The overall assessment of the song's lyrics. 
\end{itemize}

The evaluator reviewed 16 songs, with one song assigned from each of the 16 participants in the Lyrics Writing Session. As each participant produced two sets of lyrics, one set per participant was randomly selected for evaluation to reduce potential song-specific bias. The evaluation session concluded with a brief interview where evaluators provided additional insights into the strengths and weaknesses of the songs. Each evaluation session lasted approximately 30 to 40 minutes, and the evaluator was compensated with a \$10 gift card. 



\begin{figure*}
    \centering
    \includegraphics[width=\linewidth]{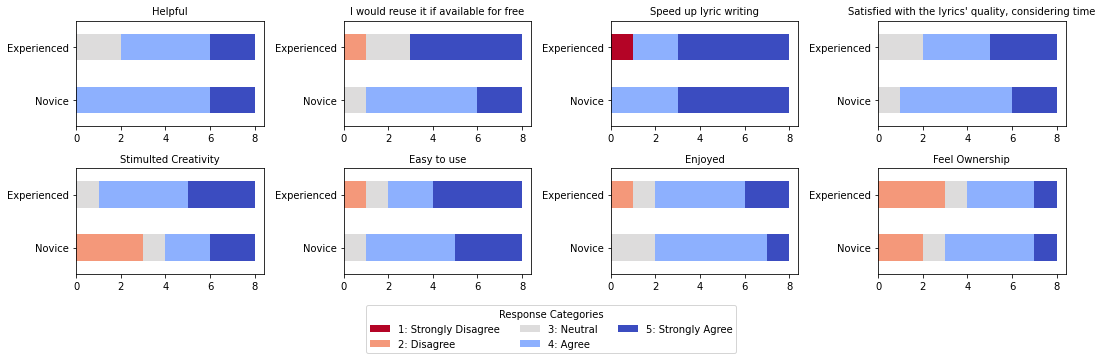}
    \caption{Questionnaire overall evaluation results comparing experienced and novice users' responses to CoLyricist. The evaluation contains 8 criteria: helpfulness, reuse intent, writing speed impact, satisfaction with output quality, creativity stimulation, ease of use, enjoyment, and sense of ownership. Responses are categorized on a 5-point Likert scale from "Strongly Disagree" to "Strongly Agree". Both user groups generally responded positively, with high agreement on the tool's helpfulness, potential for reuse, and ability to speed up lyric writing.}
    \Description{Grouped horizontal stacked bar charts showing overall questionnaire responses from experienced and novice users evaluating CoLyricist across eight criteria: helpfulness, reuse intent, writing speed impact, satisfaction with output quality, creativity stimulation, ease of use, enjoyment, and sense of ownership. Each criterion is represented by a pair of bars corresponding to experienced users and novice users. Bar segments are color-coded to indicate response categories on a five-point Likert scale, ranging from strongly disagree to strongly agree, as shown in the legend. The length of each colored segment represents the proportion of responses in that category for the corresponding criterion and user group, allowing visual comparison of response distributions between experienced and novice participants.}
    \label{fig:results_overall}
\end{figure*}


\begin{figure*}
    \centering
    \includegraphics[width=0.85\linewidth]{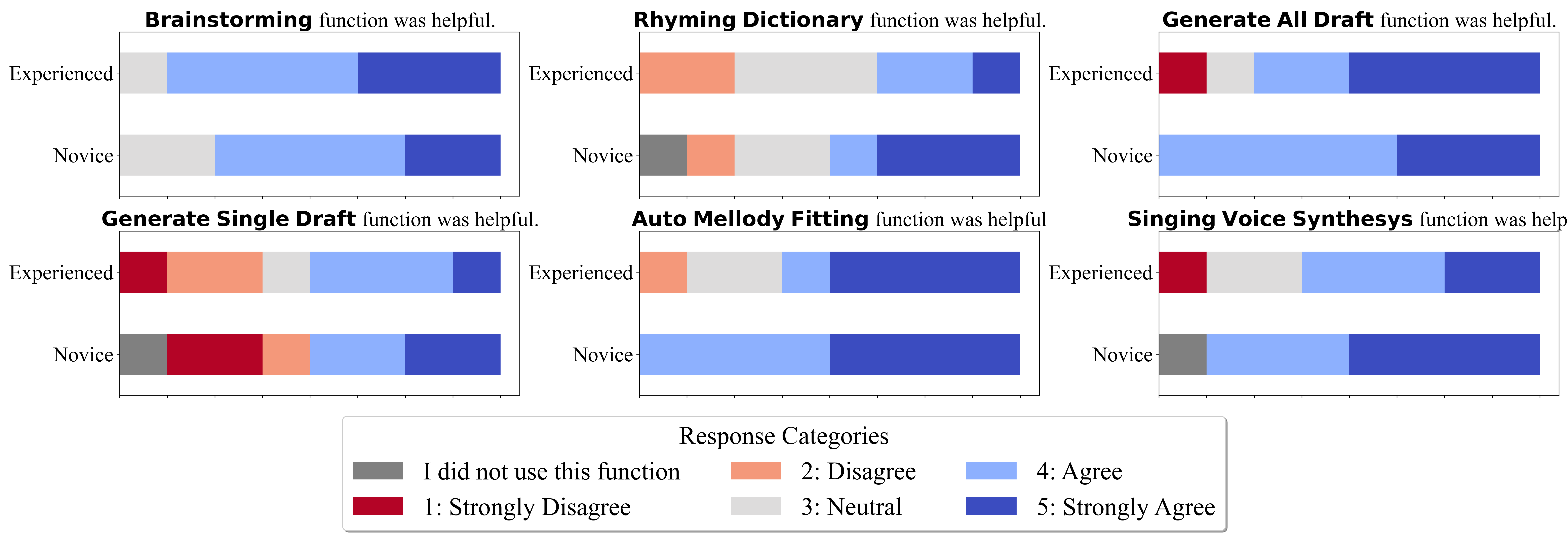}
    \caption{Questionnaire result for each feature. The chart displays ratings for six key functions of an AI-assisted lyric writing tool: Brainstorming, Rhyming Dictionary, Generate All Draft, Generate Single Draft, Auto Melody Fitting, and Singing Voice Synthesis. Responses from experienced and novice users are categorized on a 5-point Likert scale from "Strongly Disagree" to "Strongly Agree," with an additional option for "I did not use this function." The data suggests that most features were perceived as helpful, with variations between experienced and novice users in their utilization and appreciation of specific functions.}
    \Description{A set of horizontal stacked bar charts showing questionnaire responses evaluating the perceived helpfulness of six individual features of CoLyricist: Brainstorming, Rhyming Dictionary, Generate All Draft, Generate Single Draft, Auto Melody Fitting, and Singing Voice Synthesis. For each feature, two bars represent responses from experienced users and novice users. Bar segments are color-coded to indicate response categories, including an option for not using the function, followed by a five-point Likert scale ranging from strongly disagree to strongly agree, as shown in the legend. The length of each colored segment corresponds to the proportion of participants selecting that response, enabling visual comparison of feature usage and perceived helpfulness between experienced and novice users across the different functions.}
    \label{fig:results_feature_based_2}
\end{figure*}

\section{Results} 
All participants successfully completed the Lyrics Writing Session within the allocated time. In this section, we first evaluate the effectiveness of each Design Decision based on insights derived from the analysis of session recordings, questionnaires, system logs, and interview data. Following this, we discuss the quality of the lyrics created by the participants, drawing on the results from the Lyrics Assessment Session.

\subsection{Effectiveness of Workflow Aligned AI Support (D1)}
CoLyricist’s first design decisions \textbf{(D1)} was to provide AI support aligned with the typical lyric writing workflow. To evaluate the effectiveness of this design decision, we aimed to confirm two aspects: (1) whether users actually followed the workflow identified in Section \ref{sec:workflow} during the lyric writing process, and (2) whether their experience using the tool was a positive one. These two aspects are discussed in detail in the following two subsections.

\subsubsection{User Interaction with CoLyricist}
Regarding the first aspect, through qualitative observations, we found that participants used CoLyricist in alignment with the typical workflow identified during the Formative Study, confirming that the tool’s design effectively supports the natural lyric writing process. Specifically, the process followed by 12 participants typically unfolded as follows:

First, participants set the theme for the song, though the title was not always determined at this point. Next, they used the Ideation panel to collect relevant keywords. Some participants made minor adjustments to the theme or title during this stage, while others skipped this step entirely. Once enough keywords were collected, participants used the "Generate All Lyrics" feature to generate draft lyrics for the entire melody. They then moved to modifying the draft lyrics for a specific melody phrase, often using the Melody Fitting feature to adjust the lyrics to fit the melody. After finalizing a melody phrase, participants moved on to the next phrase, repeating the steps as necessary.

This step-by-step approach—finalizing one melody phrase at a time before moving to the next—was the most common workflow observed. However, four participants opted for a different approach: they first completed all the draft lyrics for the entire song to their satisfaction before moving on to finalizing the lyrics. This finding aligns with the observation from Formative Study Finding 2 (F2), suggesting that CoLyricist is flexible enough to accommodate both approaches. 

In interviews, participants who followed the all-at-once approach expressed a strong focus on ensuring the overall narrative structure of the song before refining individual phrases. Conversely, those who worked phrase-by-phrase were frequently observed returning to previous lines to ensure consistency with the rest of the song.

\subsubsection{User Satisfaction and Feedback}
Next, we analyze the second aspect: whether the experience of using CoLyricist was positive, based on participants’ feedback. The participant feedback collected in the questionnaire (see Figure 4) demonstrated overall positive reactions toward CoLyricist, especially among novice users. Below are insights drawn from these responses.

\paragraph{Overall Helpfulness} As shown in Figure 4, most participants rated CoLyricist as helpful (Strongly Agree=4, Agree=10, Neutral=2). Specifically, all novice users found it to be helpful. Also, for the question "I would reuse it if available for free", only one participant had a negative reaction (Strongly Agree=7, Agree=5, Neutral=3, Disagree=1). The user who expressed a negative opinion placed high value on the process of creating lyrics by themselves, indicating that their personal approach to lyric writing conflicted with CoLyricist's generative assistance.

\paragraph{Efficiency Improvements} Participants commonly noted that CoLyricist improved their lyric-writing speed (Strongly Agree=10, Agree=5, Strongly Disagree=1). Although most participants did not adopt the generated lyrics without modification, users frequently used the system’s suggestions as a basis for their own edits, accelerating the writing process compared to starting from scratch. Novice participants, in particular, felt the tool saved them considerable time, with one stating, \textit{"Without this tool, it would have taken me much longer, and I might not have finished the lyrics even with more time"} (P2). 

\paragraph{Quality Improvements} Regarding lyrical quality, the majority of participants were satisfied with the quality of their lyrics considering the time constraints (Strongly Agree=5, Agree=8, Neutral=3). These results suggest that CoLyricist helped maintain a certain level of quality without compromising creativity. Some users, however, mentioned that with more time, they could have further improved their lyrics. Several participants also felt that the tool stimulated creativity by providing new ideas, though this effect seemed more pronounced for experienced users. Interestingly, nearly all experienced users responded positively to the question "Stimulated Creativity" (Strongly Agree=3, Agree=4, Neutral=1), while novice users gave more mixed responses (Strongly Agree=2, Agree=2, Neutral=1, Disagree=3). The three novice users who provided negative feedback felt that the system's suggestions were too ordinary, lacking the novelty they sought. In contrast, some experienced users, while also feeling the suggestions were ordinary, appreciated how the system's outputs encouraged to come up with creative ideas of their own.

\paragraph{Process Satisfaction} In terms of ease of use, most participants found CoLyricist to be "easy to use" (Strongly Agree=7, Agree=6, Neutral=2, Disagree=1), and they reported enjoying the overall lyric writing process (Strongly Agree=3, Agree=9, Neutral=3, Disagree=1). Interviews revealed that participants appreciated the intuitive design, especially those experienced in Melody-First lyric writing. One experienced participant who consistently follows the Melody-First Style noted, \textit{"It aligns perfectly with my process, so it was very easy to use"} (P13). On the other hand, the participant who rated "Easy to use" as Disagree mentioned that they usually make slight adjustments to the melody during the lyric writing process, but the fixed melody in the experiment limited their flexibility. Several participants also praised the interactive nature of CoLyricist, specifically highlighting the presence of "Draft Lyrics," which increased the level of interactivity by allowing them to influence the generated lyrics, rather than simply generating final lyrics directly from a theme. This confirms the effectiveness of Design Decision 1 in supporting users' processes.

\paragraph{Ownership Concerns} A subset of participants expressed concerns about feeling a lack of ownership over their work (Strongly Agree=2, Agree=7, Neutral=2, Disagree=5). Among those who disagreed, 3 participants noted that with more time, they would have felt more ownership over the lyrics. One experienced participant, who also selected "Disagree," made an interesting remark: \textit{"I don’t always feel complete ownership over songs even when collaborating with other songwriter, so this is not a major issue for me."} \textbf{(P15)} This indicates that reduced ownership is not always perceived as a negative aspect. On the other hand, one novice participant who did feel ownership mentioned that as long as they had full control over the lyrics draft, they felt ownership, as it allowed them to effectively convey their message in the final lyrics.
\vspace{10pt}

In summary, both experienced and novice participants generally followed the workflow we had anticipated, and overall, they perceived the experience positively. This confirms the effectiveness of our first Design Decision (D1). Moreover, they demonstrate that while CoLyricist was designed with the workflows of experienced lyricists, its features are effective in supporting novice users as well.

\subsection{User Usage of AI Support (D2-D4)}
The remaining three Design Decisions (D2–D4) focus on the design of AI support provided at each step of the lyric writing process. Figure 5 summarizes user evaluations of the various features. In the following three subsections, we discuss the effectiveness of D2, D3, and D4, drawing on the insights from Figure 5 and participants' interview responses.

\subsubsection{Feedback on Ideation Features (D2)}
Among the ideation features, the \textbf{Brainstorming feature} was perceived as particularly valuable. This feature received the most praise from experienced users (Strongly Agree=3, Agree=4, Neutral=1). These participants relied heavily on the early stages of ideation, using the Brainstorming function to deepen their ideas. Notably, experienced users employed an iterative approach, repeatedly inputting outputs from the tool to further expand on specific phrases. This behavior highlights the feature's potential for enhancing creative exploration in the ideation phase. Specifically, an experienced participants quoted \textit{"The brainstorming tools were useful for coming up with ideas and supporting any themes that I already had"} (P10).

In contrast, the \textbf{Rhyming Recommendation Feature} received mixed reviews, with some participants pointing out quality issues. One experienced user remarked, \textit{"Helpful as it does its job as a rhyme finder. But sometimes when I'm not satisfied with the current rhymes, the regenerated results show nothing new. An indicator of rhymes that's not so common would be helpful"} (P9). Although this feature was designed to propose words closely related to the theme (C2), this focus may have inadvertently reduced the diversity of the suggestions. Nonetheless, while this feedback highlights the need for improvement in the quality, it also underscores the demand for the feature.
These feedback on the two features strongly support the validity of our design decisions (D2).





\subsubsection{Feedback on Draft Lyrics Generation Features (D3)}
The Draft Lyrics generation features yielded varied reactions among participants. The \textbf{"Generate All Draft" feature} was well-received, with participants appreciating its ability to provide a solid starting point for the song. This feature saved time by offering an initial framework, as one participant noted, \textit{"It is handy for sketching out a draft or overall song structure"} (P6). This feedback strongly supports our design decision (D3). However, a participant who "Strongly Disagreed" with the helpfulness of this feature expressed concerns about ownership, stating, \textit{"It just made me feel like I wasn't actually songwriting when I used full draft generation feature"} (P10).

In contrast, the \textbf{"Generate Single Draft" feature} received more mixed feedback. While some participants found it useful, many noted that the generated suggestions often lacked coherence with the surrounding lyrics, making them challenging to incorporate. Despite these concerns, system logs revealed frequent usage of this feature, averaging around six times per session. This suggests a demand for the feature, provided its quality can be improved, further supporting our design decision (D3).



\subsubsection{Feedback on Melody-Fitting Features (D4)}
\label{sec:result_melody_fitting}
The \textbf{Melody-Fitting feature} was well-received by both groups, particularly novice users. All novice participants rated the feature as highly helpful (Strongly Agree=4, Agree=4), with half identifying it as the most helpful feature overall during interviews. One participant noted, \textit{"The Fit to Melody function was very helpful. It significantly speeds up my work"} (P4).
Experienced users were slightly less enthusiastic about this feature (Strongly Agree=4, Agree=1, Neutral=2, Disagree=1), citing concerns about the quality of AI-generated suggestions. While they found the feature helpful, the outputs sometimes required some revisions to align with their creative intent. As one experienced participant mentioned, \textit{"The melody-fitting function was the most important feature that I would use in my actual work. Although it contained some grammatical errors, I found it easy to revise"} (P16).

Interestingly, during each lyric-writing session consisting of 8 melody phrases, novice participants adopted the Melody-Fitting suggestions without any modifications for an average of 5.19 phrases, compared to 2.79 phrases for experienced participants. This discrepancy highlights that experienced users were more committed to ensuring the lyrics reflected their personal creative intent.

The \textbf{Voice Synthesis feature} revealed an interesting contrast: novice users rated it more favorably, but system logs showed that experienced users utilized it more frequently. On average, experienced participants used Voice Synthesis 15.6 times per session compared to 9.5 times for novices. This suggests that experienced users placed greater emphasis on fine-tuning the alignment between lyrics and melody. Conversely, a novice participant noted \textit{"I primarily focused on the message of the lyrics, paying less attention to how well the words fit the melody, as long as the syllable count matched" (P6).}
In summary, both the Melody-Fitting and Voice Synthesis features were positively perceived by participants, confirming the validity of our design decision (D4).

\begin{figure}
    \centering
    \includegraphics[width=1.0\linewidth]{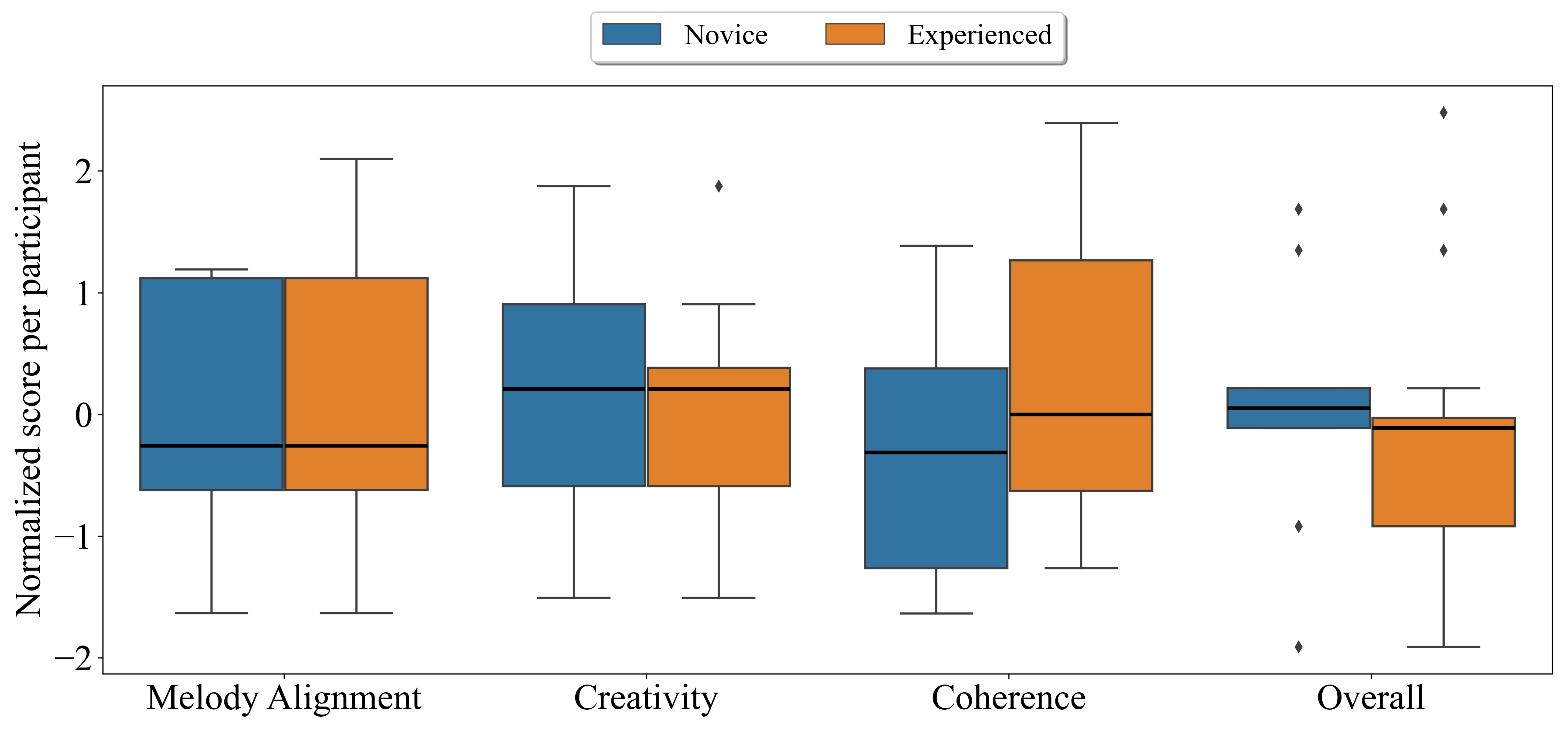}
    \caption{User Created Lyrics Assessment results. The box plot displays normalized scores for four criteria: Melody Alignment, Creativity, Coherence, and Overall quality. The result suggests variations in performance between novice and experienced users across the different assessment criteria is not significant.}
    \Description{Box plots showing normalized assessment scores of user-created lyrics for four evaluation criteria: Melody Alignment, Creativity, Coherence, and Overall quality. For each criterion, two box plots represent scores from novice users and experienced users, as indicated by the legend. The vertical axis shows normalized scores per participant, while each box depicts the distribution of scores, including the median, interquartile range, whiskers, and outliers. This visualization allows comparison of score distributions between novice and experienced participants across the four assessment criteria.}
    \label{fig:results_lyrics_assessment}
\end{figure}

\subsection{Expert Assessment}

As described in Section \ref{sec:assessment-setting}, we invited two professional experts with strong musical backgrounds for the evaluation of created lyrics in the user study. One evaluator is a professional producer (E1), while the other has over 10 years of experience in instrumental performance, more than 6 years of vocal performance, and experience in songwriting as a hobby (E2). The average scores assigned by the evaluators were 3.06 (E1) and 2.81 (E2), with an inter-annotator agreement rate of 0.44 (measured by Pearson correlation coefficient).

\subsubsection{Narrowing the Skill Gap Between Novices and Experts}
Figure \ref{fig:results_lyrics_assessment} illustrates the results, showing the normalized scores for each metric across all songs and evaluators. 
Scores were normalized per evaluator to account for differences in how each expert used the five-point rating scale (e.g., stricter or more lenient average ratings).
The results indicate that, given the time constraints, the quality of work produced by novice and experienced writers was broadly comparable.

First, the melody alignment scores showed similar distributions for both groups, suggesting that the tool effectively helps novice writers achieve alignment comparable to that of experienced writers. An independent two-sample t-test on melody alignment scores yielded a p-value of 0.94, indicating no significant difference. Novice writers scored slightly lower on lyrical coherence, which is an expected outcome given their lack of experience.
Scores for overall quality also showed comparable distributions. A t-test yielded a p-value of 0.62, indicating no significant difference here either.

However, while the difference was not statistically significant, the median score for experienced participants was slightly lower. This may be related to the fact that three experienced participants mentioned during interviews that they appreciated having more time to polish their lyrics, even though they managed to complete them in time. This aligns with observations that experienced participants tended to revise system suggestions more extensively (Section \ref{sec:result_melody_fitting}).
These findings suggest that CoLyricist successfully narrows the gap between novice and experienced participants, particularly in technical aspects such as melody alignment.

A potential concern is the possibility that CoLyricist might inadvertently constrain the creativity of experienced users. Studies on writing assistance tools have reported instances of homogenization \cite{Padmakumar2023-dg}. In this study, however, many experienced participants found CoLyricist "helpful" and it “stimulated their creativity” (Figure \ref{fig:results_overall}). However, the possibility that their creativity was subconsciously limited cannot be entirely ruled out. Future work should include evaluations of lyrics written without the system to compare and assess the impact of the tool on creativity more rigorously.

\subsection{Summary} 
Overall, participants responded positively to \projectname, and the design effectively supported lyricists' comprehensive creative workflows seamlessly, for both experienced and novice participants. Although some quality improvements to the tool's generative features are needed, the tool's design decisions were validated. Additionally, novice users and experienced users found different features most helpful, with novices relying on the Melody-Fitting feature and experienced users valuing the Brainstorm function. 
Furthermore, with the aid of \projectname, novice participants produced work of comparable quality to experienced writers in terms of melody alignment and overall quality, suggesting the tool effectively narrows the gap between skill levels in technical aspects of songwriting.

\section{Discussions and Future Work}
This section discusses the key challenges identified in our study and outlines future directions for the development of lyric-writing support tools. We begin in Section ~\ref{sec:discussion_limitation} by addressing the limitations of our current work. 
Next, in Sections ~\ref{sec:discussion_novice_challenge} and ~\ref{sec:discussion_update_features}, we focus on the design implications for improving CoLyricist. Section ~\ref{sec:discussion_novice_challenge} analyzes the differences in how novice and experienced users interact with the tool, while Section ~\ref{sec:discussion_update_features} builds on these findings to explore necessary interaction design elements for enhancing usability across user groups.
Finally, Sections ~\ref{sec:discussion_optimal_workflow} turn to broader discussions about the workflow itself, which served as the foundation for CoLyricist’s design. We examine how the defined workflow can be further refined to better reflect a user-centered design perspective. 
 
\subsection{Limitations}
\label{sec:discussion_limitation}
This work has several limitations that suggest directions for future research. First, the workflows and stage-specific challenges identified in this study were derived from interviews with experienced hobbyist lyricists and may not generalize to professional lyricists, whose creative processes and constraints often differ, for example due to time pressure and higher songwriting proficiency~\cite{baskerville2013music, Dodson2024, Kardzha2024-vx}. Future studies should examine whether similar workflows and challenges emerge among professional lyricists. 

Second, the experimental setting of the user study did not allow participants to edit the melody. While some participants reported wanting to modify the melody during lyric writing, CoLyricist currently assumes a fixed melody, which affected usability in such cases. In real-world songwriting, lyric and melody creation are often interdependent, with creators iterating between the two using multiple tools. Supporting melody editing directly or enabling smoother integration with external composition tools remains an important direction for future work, informed by prior research on melody composition support~\cite{Louie2020-um, Kobayashi2024-jx, Larsen-ab}.

Finally, the current generation model used in CoLyricist does not explicitly incorporate higher-level musical structure (e.g., verse–\allowbreak chorus relationships). While considering such structural information would likely improve generation quality, the primary focus of this study was on evaluating the user experience and interaction design of a workflow-aligned lyric-writing tool rather than optimizing model output performance. Incorporating musical structure into the generation model remains an important direction for future work.

\begin{figure*}[t]
    \centering
    \begin{subfigure}{0.49\textwidth}
        \centering
        \includegraphics[width=\linewidth]{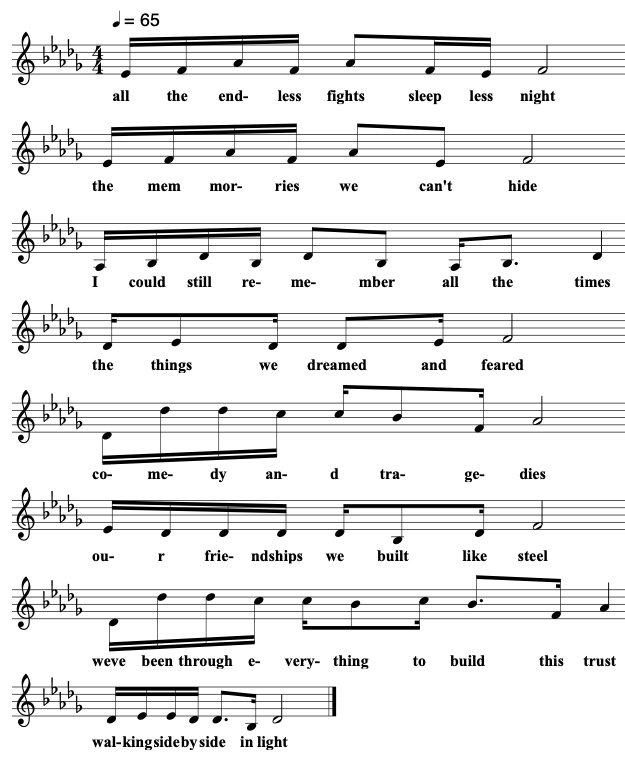}
        \caption{Lyrics Created by experienced participants}
        \label{fig:lyrics_example_good}
    \end{subfigure}
    \hfill
    \begin{subfigure}{0.49\textwidth}
        \centering
        \includegraphics[width=\linewidth]{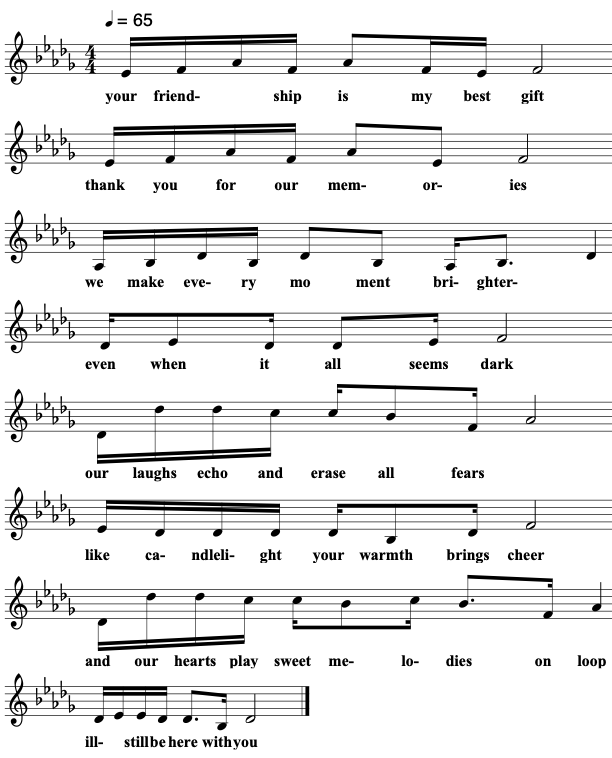}
        \caption{Lyrics Created by Novice participants}
        \label{fig:lyrics_example_bad}
    \end{subfigure}
    \caption{Examples of lyrics showing differences in melody alignment. (a) Lyrics by an experienced participant with high scores (E1: Good, E2: Good), demonstrating strong prosody and effective rhyming. (b) Lyrics by a novice participant with lower scores (E1: Poor, E2: Very Poor), featuring misaligned key words and mismatched syllable counts.}
    \Description{Side-by-side musical score excerpts with lyrics illustrating examples of melody-aligned lyric writing by different participants. The left panel shows lyrics created by an experienced participant, where syllables are closely aligned with note durations and melodic structure across multiple phrases. The right panel shows lyrics created by a novice participant, where syllable placement and word segmentation differ across notes, resulting in less consistent alignment with the melody. Each excerpt presents staff notation with tempo markings and lyrics placed beneath the notes, allowing visual comparison of how lyrical content is mapped onto the same melodic framework by experienced and novice users.}
    \label{fig:lyrics_example}
\end{figure*}

\subsection{Challenges in Editing Lyrics for Novices}
\label{sec:discussion_novice_challenge}
Figure \ref{fig:lyrics_example} (a) showcases lyrics written by an experienced participant that received high evaluations specifically for "Melody Alignment" (E1: Good, E2: Good) in the expert assessment session. Regarding melody alignment, E1 noted, \textit{"The prosody matches the melody perfectly, making the lyrics feel very natural."} Similar to speaking, when singing, singers emphasize important words in lyrics (e.g., verbs or nouns). These important words should align with prominent notes in melody for good prosody and naturalness. \cite{Zhao2024-ic}. For example, in the first line, the important words "fight" and "nights" align seamlessly with prominent notes, emphasizing them through the melody.
The lyrics also feature effective rhyming, such as [night, hide, times] and [tragedies, steel]. E2 commented, \textit{"Proper rhyming significantly influences the overall impression."} However, the participant achieved this rhyming by manually editing the system's suggestions.

On the other hand, Figure \ref{fig:lyrics_example} (b) shows lyrics written by a novice participant that received lower score specifically for "Melody Alignment" ( E1: Poor, E2: Very Poor). The melody alignment here has several issues. For instance, in forth and fifth line, the important words ``seems'',``laughs'', and ``echo'' are all assigned to non-prominent notes, creating a sense of awkwardness. Additionally, syllable counts in words like ``erase'' (line 5) and ``candlelight'' (line 6) do not match the number of notes (have more syllables than corresponding notes), making these words difficult to sing and harder for listeners to perceive. Notably, these issues often occurred at the lines where edited by the novice participant themselves, suggesting that editing system-generated lyrics remains a challenge for novices. This aligns with the observation discussed in Section \ref{sec:result_melody_fitting}, where novices tend to use the Melody-Fitting suggestions without making modifications. In fact, when dissatisfied with the Melody-Fitting suggestions, novice participants were often observed revising the Draft Lyrics instead of directly editing the lyrics and then reapplying the Melody-Fitting feature for new suggestions.

Taken together, these observations suggest that lyric-writing support tools should not only generate melody-aligned lyrics, but also provide explicit guidance for editing and refinement, particularly for novice users. Designing interactions that help novices understand why certain edits disrupt prosody or singability may be critical.

\subsection{Enhancing Control and Interactivity in CoLyricist}
\label{sec:discussion_update_features}

In the user study, one experienced participant noted: \textit{"When I had to edit the (generated) phrases, I sometimes felt like I was fighting with the AI tool rather than being assisted by it"} (P10). This highlights challenges in aligning AI outputs with user intent. In this subsection, we discuss how to improve flexibility, interactivity, and personalization to address these issues and support more seamless lyric writing.

\subsubsection{Flexible Control for Melody Fitting}
Several participants requested the ability to specify rhyming positions and phonemes during Melody-Fitting. While the tool enables rhyming word searches, positioning them with contextual appropriateness remains challenging. Others also asked for functionality to lock specific words to certain notes, such as fixing hook words to key melodic positions. Currently, CoLyricist supports such constraints only at the phrase level, highlighting the need for more fine-grained control.

Although each rule may be simple, satisfying multiple constraints simultaneously while considering melodic structure is not trivial. \citet{Zhang2024-nq} proposed a constrained decoding technique using Deterministic Finite Automata (DFA) to enforce arbitrary constraints in LLM generation. Representing such rules with DFA may offer a path toward satisfying these constraints reliably.

Beyond these control requests, users highlighted the absence of style-specific features for customizing tone, genre, or linguistic style. For instance, \citet{Laban2024-ya} developed a document editing tool that enables users to direct style through natural language chat or by selecting from predefined style presets. While experienced users may articulate their stylistic preferences clearly, novices might benefit more from preset-based interactions. Exploring such options remains an area for future work.

\subsubsection{Interactive AI Guidance}
As discussed in Section~\ref{sec:discussion_novice_challenge}, novice users often struggled to refine generated lyrics due to limited experience. A promising direction is to add interactive guidance features that suggest directions for improving prosody or phrasing, rather than simply offering full-line replacements.

In non-lyric writing tools, many systems provide automatic feedback to guide revision~\cite{Karolus2023-ia, Wambsganss2022-ym, Afrin2021-up, Benharrak2024-ov}. For example, \citet{Laban2024-ya} offer interactive reviews through chat, while \citet{Karolus2023-ia} focus on readability evaluation and feedback, which has been shown to improve both users' self-awareness and writing skills.

However, lyrics involve unique evaluation criteria such as singability and rhyming, which depend on melodic alignment as well as semantics. Therefore, it is insufficient to adopt existing text review tools directly. Instead, there is a need for LLM-based systems that generate feedback based on known criteria for lyric quality~\cite{Nichols2009-tp, Zhao2024-ic}, such as singability and prosodic alignment. Such features could not only make the tool more intuitive but also promote long-term improvement of users' songwriting abilities.

Additionally, because lyric writing is a highly artistic task with wide personal variation, feedback should also reflect individual preferences. This aspect is discussed in the next subsection.

\subsubsection{Towards Personalized Suggestions}
As observed in the user study, participants had varying reactions to generated content—some found it mundane, while others appreciated its inspirational value. This variance likely stems from differences in personal preferences. For instance, one participant remarked, \textit{"I write lyrics based on personal experiences and memories. The system’s suggestions didn’t help trigger my own memories"} (P14). To address this, the system must better understand individual user contexts to offer more meaningful and personalized outputs.

CoLyricist's Melody-Fitting interface offers a way to collect implicit preference signals—when users select among alternative lyric suggestions, their choices reveal preferences without requiring explicit input. These signals, if effectively leveraged with personalization techniques~\cite{Rafailov2023-et, Christiano2017-je, Koyama2022-kb}, could enable more adaptive and creative support. However, collecting sufficient implicit data per user remains challenging, especially in creative tasks with high variation. To complement this, incorporating explicit preference signals—such as asking users to state their goals, stylistic preferences, or target emotional tone in a brief pre-creation session—could further enhance personalization. 

In fact, several studies in non-lyric domains have explored leveraging user preferences for personalization \cite{Liu2025-sz}. For example, \citet{Zhou2021-cb} proposed a method that uses implicit preference data—such as user selections among generated options—to guide Bayesian optimization in the latent space of a VAE, gradually personalizing outputs. However, this approach cannot be directly applied to Transformer-based LLMs. In contrast, \citet{Li2024-kr} demonstrated a method that integrates both implicit choices and explicit user feedback to tailor generative outputs. While promising, their work focused on conversational data, which is easier to collect and still required extensive annotation. Applying these techniques to lyric writing would demand careful interaction design and modeling.

\subsection{Exploring Better Workflows for Lyrics Writing Support Tool}
\label{sec:discussion_optimal_workflow}

As discussed, aligning CoLyricist's design with the high-level workflows identified in the Formative Study proved effective. However, observations of participants' usage revealed deviations from the anticipated workflows, especially at a finer level of interaction. For example, with the "Generate Draft Lyrics" feature (D3), we originally expected participants to use the "Generate Single Draft" function to complete only difficult lines after drafting easier ones themselves, as suggested by Challenge (C3) from the Formative Study. Contrary to this assumption, many participants first used the "Generate All Draft" function to quickly populate the entire song framework, and then relied on "Generate Single Draft" for refinements.

This behavior may have been influenced by the limited time available in the User Study, encouraging efficiency. However, it also raises the possibility that this interaction pattern represents an optimal workflow unique to human-AI co-creation, where users employ AI to create a rough foundation and refine it iteratively.

In creative support tools, including those for lyric writing, the balance between automation by AI and manual control by the user plays a significant role in shaping user satisfaction and their sense of ownership~\cite{Biermann2022-jk, Xu2024-dj, Haase2024-ql}. In the case of CoLyricist, participants generally reported positive experiences, suggesting that fully automatic drafting may contribute to an enhanced lyric-writing experience.

In particular, \citet{Xu2024-dj} identify two key components of ownership in creative tasks: the perceived originality of the work and the user's level of contribution. Originality is enhanced when users can infuse personal ideas, while the level of contribution relates to the sense of control during the creative process. We hypothesize that even with AI-generated drafts, CoLyricist preserved users' sense of originality and contribution through its interaction design, which allows users to convey detailed intent to the system. Furthermore, because users could freely edit draft outputs and use the Melody Fitting feature to choose from multiple line-level suggestions, they remained actively involved in the creative process—reinforcing their sense of contribution.

While CoLyricist provides support aligned with typical lyric-writing workflows, a deeper understanding of which stages users prefer to automate versus control manually may reveal alternative workflows. These insights could lead to designs that selectively skip certain steps or even support entirely new workflows tailored to individual preferences and creative styles.




\section{Conclusion}
In this paper, we presented CoLyricist, an AI-assisted lyric writing tool designed to align with typical lyric writing workflows. Based on interviews with 10 amateur songwriters, we identified key stages of the lyric-writing process—Theme Setting, Ideation, Lyrics Drafting, and Melody Fitting—and developed CoLyricist to provide AI support for each stage. Our user study with 16 participants, including both novice and experienced lyricists, showed that CoLyricist effectively enhances the songwriting process. Novice users found the Melody-Fitting feature particularly helpful, while experienced users valued the Ideation support. Overall, participants reported increased productivity and satisfaction. Furthermore, expert reviews of the lyrics written during the user study indicated no significant difference in quality between those written by novice and experienced users, suggesting that CoLyricist narrows skill gap between novice and experienced lyricists. CoLyricist shows promise as a valuable tool for supporting lyricists and can evolve to become even more effective with further development.

\section{Generate AI Use Disclosure}

This paper content was written with the assistance of generative AI tools, such as OpenAI GPT-5.
These tools were used to help improve the clarity and readability of the text. The authors reviewed and edited the content as needed and take full responsibility for the final version of the paper.


\bibliographystyle{ACM-Reference-Format}
\bibliography{ref_main}

@INPROCEEDINGS{Ma2021-ga,
  title     = "{AI-Lyricist}: Generating Music and Vocabulary Constrained
               Lyrics",
  booktitle = "Proceedings of the 29th {ACM} International Conference on
               Multimedia",
  author    = "Ma, Xichu and Wang, Ye and Kan, Min-Yen and Lee, Wee Sun",
  publisher = "Association for Computing Machinery",
  pages     = "1002--1011",
  series    = "MM '21",
  month     =  oct,
  year      =  2021,
  address   = "New York, NY, USA",
  location  = "Virtual Event, China"
}

@ARTICLE{Ding2024-gn,
  title         = "{SongComposer}: A Large Language Model for Lyric and Melody
                   Composition in Song Generation",
  author        = "Ding, Shuangrui and Liu, Zihan and Dong, Xiaoyi and Zhang,
                   Pan and Qian, Rui and He, Conghui and Lin, Dahua and Wang,
                   Jiaqi",
  month         =  feb,
  year          =  2024,
  archivePrefix = "arXiv",
  primaryClass  = "cs.SD",
  eprint        = "2402.17645"
}

@ARTICLE{Oliveira2015-yc,
  title    = "{Tra-la-Lyrics} 2.0: Automatic Generation of Song Lyrics on a
              Semantic Domain",
  author   = "Oliveira, Hugo Gon{\c c}alo",
  journal  = "Journal of Artificial General Intelligence",
  volume   =  6,
  number   =  1,
  pages    = "87--110",
  month    =  dec,
  year     =  2015
}

@INPROCEEDINGS{Watanabe2017-as,
  title     = "{LyriSys}: An Interactive Support System for Writing Lyrics
               Based on Topic Transition",
  booktitle = "Proceedings of the 22nd International Conference on Intelligent
               User Interfaces",
  author    = "Watanabe, Kento and Matsubayashi, Yuichiroh and Inui, Kentaro
               and Nakano, Tomoyasu and Fukayama, Satoru and Goto, Masataka",
  publisher = "Association for Computing Machinery",
  pages     = "559--563",
  series    = "IUI '17",
  month     =  mar,
  year      =  2017,
  address   = "New York, NY, USA",
  location  = "Limassol, Cyprus"
}

@ARTICLE{Yamashita2022-pw,
  title     = "Implementation and evaluation of a collaborative lyric-writing
               support system using a lyric association map",
  author    = "Yamashita, Meguru and Sato, Kiwamu and Doi, Akio",
  journal   = "Multimodal Technol. Interact.",
  publisher = "MDPI AG",
  volume    =  6,
  number    =  4,
  pages     =  23,
  month     =  apr,
  year      =  2022,
  language  = "en"
}

@ARTICLE{Abe2012-zt,
  title   = "A Japanese lyrics writing support system for amateur songwriters",
  author  = "Abe, Chihiro and Ito, Akinori",
  journal = "Proceedings of The 2012 Asia Pacific Signal and Information
             Processing Association Annual Summit and Conference",
  pages   = "1--4",
  month   =  dec,
  year    =  2012
}

@INPROCEEDINGS{Goncalo_Oliveira2017-lb,
  title     = "{Co-PoeTryMe}: a {Co-Creative} Interface for the Composition of
               Poetry",
  booktitle = "Proceedings of the 10th International Conference on Natural
               Language Generation",
  author    = "Gon{\c c}alo Oliveira, Hugo and Mendes, Tiago and Boavida, Ana",
  editor    = "Alonso, Jose M and Bugar{\'\i}n, Alberto and Reiter, Ehud",
  publisher = "Association for Computational Linguistics",
  pages     = "70--71",
  month     =  sep,
  year      =  2017,
  address   = "Santiago de Compostela, Spain"
}

@ARTICLE{Zhu2023-qe,
  title         = "A Survey of {AI} Music Generation Tools and Models",
  author        = "Zhu, Yueyue and Baca, Jared and Rekabdar, Banafsheh and
                   Rawassizadeh, Reza",
  month         =  aug,
  year          =  2023,
  archivePrefix = "arXiv",
  primaryClass  = "cs.SD",
  eprint        = "2308.12982"
}

@INPROCEEDINGS{Nichols2009-tp,
  title     = "Relationships Between Lyrics and Melody in Popular Music",
  booktitle = "Proceedings of the 10th International Society for Music
               Information Retrieval Conference, {ISMIR} 2009, Kobe
               International Conference Center, Kobe, Japan, October 26-30,
               2009",
  author    = "Nichols, Eric Paul and Morris, Dan and Basu, Sumit and Raphael,
               Christopher",
  publisher = "unknown",
  pages     = "471--476",
  month     =  jan,
  year      =  2009
}

@INPROCEEDINGS{Watanabe2018-rt,
  title     = "A {Melody-Conditioned} Lyrics Language Model",
  booktitle = "Proceedings of the 2018 Conference of the North {A}merican
               Chapter of the Association for Computational Linguistics: Human
               Language Technologies, Volume 1 (Long Papers)",
  author    = "Watanabe, Kento and Matsubayashi, Yuichiroh and Fukayama, Satoru
               and Goto, Masataka and Inui, Kentaro and Nakano, Tomoyasu",
  editor    = "Walker, Marilyn and Ji, Heng and Stent, Amanda",
  publisher = "Association for Computational Linguistics",
  pages     = "163--172",
  month     =  jun,
  year      =  2018,
  address   = "New Orleans, Louisiana"
}

@INPROCEEDINGS{Sheng2021-nq,
  title     = "{SongMASS}: Automatic Song Writing with Pre-training and
               Alignment Constraint",
  booktitle = "Proceedings of the {AAAI} Conference on Artificial Intelligence",
  author    = "Sheng, Zhonghao and Song, Kaitao and Tan, Xu and Ren, Yi and Ye,
               Wei and Zhang, Shikun and Qin, Tao",
  month     =  may,
  year      =  2021
}

@ARTICLE{Knosche2005-perception,
  title         = "Perception of phrase structure in music.",
  author        = "Knösche, TR. and Neuhaus, C. and  Haueisen, J and Alter, K. and Maess, B and Witte, OW and Friederici, AD",
  month         =  Apr,
  year          =  2005,
  journal = "Hum Brain Mapp.",
  doi  = "doi: 10.1002/hbm.20088. PMID: 15678484"
}

@INPROCEEDINGS{Tian2023-di,
  title     = "Unsupervised {Melody-to-Lyrics} Generation",
  booktitle = "Proceedings of the 61st Annual Meeting of the Association for
               Computational Linguistics (Volume 1: Long Papers)",
  author    = "Tian, Yufei and Narayan-Chen, Anjali and Oraby, Shereen and
               Cervone, Alessandra and Sigurdsson, Gunnar and Tao, Chenyang and
               Zhao, Wenbo and Chung, Tagyoung and Huang, Jing and Peng, Nanyun",
  editor    = "Rogers, Anna and Boyd-Graber, Jordan and Okazaki, Naoaki",
  publisher = "Association for Computational Linguistics",
  pages     = "9235--9254",
  month     =  jul,
  year      =  2023,
  address   = "Toronto, Canada"
}

@inproceedings{Lee2024-hb,
author = {Lee, Mina and Gero, Katy Ilonka and Chung, John Joon Young and Shum, Simon Buckingham and Raheja, Vipul and Shen, Hua and Venugopalan, Subhashini and Wambsganss, Thiemo and Zhou, David and Alghamdi, Emad A. and August, Tal and Bhat, Avinash and Choksi, Madiha Zahrah and Dutta, Senjuti and Guo, Jin L.C. and Hoque, Md Naimul and Kim, Yewon and Knight, Simon and Neshaei, Seyed Parsa and Shibani, Antonette and Shrivastava, Disha and Shroff, Lila and Sergeyuk, Agnia and Stark, Jessi and Sterman, Sarah and Wang, Sitong and Bosselut, Antoine and Buschek, Daniel and Chang, Joseph Chee and Chen, Sherol and Kreminski, Max and Park, Joonsuk and Pea, Roy and Rho, Eugenia Ha Rim and Shen, Zejiang and Siangliulue, Pao},
title = {A Design Space for Intelligent and Interactive Writing Assistants},
year = {2024},
isbn = {9798400703300},
publisher = {Association for Computing Machinery},
address = {New York, NY, USA},
url = {https://doi.org/10.1145/3613904.3642697},
doi = {10.1145/3613904.3642697},
booktitle = {Proceedings of the 2024 CHI Conference on Human Factors in Computing Systems},
articleno = {1054},
numpages = {35},
location = {Honolulu, HI, USA},
series = {CHI '24}
}

@INPROCEEDINGS{Yuan2022-ds,
  title     = "Wordcraft: Story Writing With Large Language Models",
  booktitle = "27th International Conference on Intelligent User Interfaces",
  author    = "Yuan, Ann and Coenen, Andy and Reif, Emily and Ippolito, Daphne",
  publisher = "Association for Computing Machinery",
  pages     = "841--852",
  series    = "IUI '22",
  month     =  mar,
  year      =  2022,
  address   = "New York, NY, USA",
  location  = "Helsinki, Finland"
}

@INPROCEEDINGS{Wu2022-nz,
  title     = "{AI} Chains: Transparent and Controllable Human-{AI} Interaction
               by Chaining Large Language Model Prompts",
  author    = "Wu, Tongshuang and Terry, Michael and Cai, Carrie Jun",
  booktitle = "Proceedings of the 2022 CHI Conference on Human Factors in
               Computing Systems",
  publisher = "Association for Computing Machinery",
  address   = "New York, NY, USA",
  number    = "Article 385",
  pages     = "1--22",
  series    = "CHI '22",
  month     =  apr,
  year      =  2022
}

@inproceedings{Lee2022-mr,
author = {Lee, Mina and Liang, Percy and Yang, Qian},
title = {CoAuthor: Designing a Human-AI Collaborative Writing Dataset for Exploring Language Model Capabilities},
year = {2022},
isbn = {9781450391573},
publisher = {Association for Computing Machinery},
address = {New York, NY, USA},
url = {https://doi.org/10.1145/3491102.3502030},
doi = {10.1145/3491102.3502030},
booktitle = {Proceedings of the 2022 CHI Conference on Human Factors in Computing Systems},
articleno = {388},
numpages = {19},
location = {New Orleans, LA, USA},
series = {CHI '22}
}

@INPROCEEDINGS{Settles2010-wm,
  title     = "Computational Creativity Tools for Songwriters",
  author    = "Settles, Burr",
  booktitle = "Proceedings of the NAACL HLT 2010 Second Workshop on
               Computational Approaches to Linguistic Creativity",
  pages     = "49--57",
  year      =  2010
}

@misc{masterwriter,
  author = {MasterWriter Inc.},
  title = {MasterWriter},
  year = {2024},
  url = {https://masterwriter.com/},
  lastaccessed = {August 2, 2024}
}

@misc{lazyjot,
  author = {Lazyjot Team},
  title = {Lazyjot},
  year = {2024},
  url = {https://lazyjot.com/},
  lastaccessed = {August 2, 2024}
}

@misc{AISongLyricGenerator,
  author = {Writecream Team},
  title = {{A}{I} {S}ong {L}yric {G}enerator with {M}elody --- Writecream},
  year = {2024},
  url = {https://www.writecream.com/ai-song-lyric-generator-with-melody/},
  lastaccessed = {August 2, 2024}
}

@inproceedings{Rongsheng2020Youling,
    title = "Youling: an {AI}-assisted Lyrics Creation System",
    author = "Zhang, Rongsheng  and
      Mao, Xiaoxi  and
      Li, Le  and
      Jiang, Lin  and
      Chen, Lin  and
      Hu, Zhiwei  and
      Xi, Yadong  and
      Fan, Changjie  and
      Huang, Minlie",
    editor = "Liu, Qun  and
      Schlangen, David",
    booktitle = "Proceedings of the 2020 Conference on Empirical Methods in Natural Language Processing: System Demonstrations",
    month = oct,
    year = "2020",
    address = "Online",
    publisher = "Association for Computational Linguistics",
    url = "https://aclanthology.org/2020.emnlp-demos.12/",
    doi = "10.18653/v1/2020.emnlp-demos.12",
    pages = "85--91"
}

@inproceedings{Qian2022Training,
  title={Training strategies for automatic song writing: A unified framework perspective},
  author={Qian, Tao and Shi, Jiatong and Guo, Shuai and Wu, Peter and Jin, Qin},
  booktitle={ICASSP 2022-2022 IEEE International Conference on Acoustics, Speech and Signal Processing (ICASSP)},
  pages={4738--4742},
  year={2022},
  organization={IEEE}
}

@misc{openai_gpt3_5,
  title = {GPT-3.5},
  author = {{OpenAI}},
  year = {2021},
  howpublished = {\url{https://platform.openai.com/docs/models/gpt-3.5}},
  note = {Accessed: 2024-05-22}
}

@article{Ricardo2020Yake,
  title={YAKE! Keyword extraction from single documents using multiple local features},
  author={Campos, Ricardo and Mangaravite, V{\'\i}tor and Pasquali, Arian and Jorge, Al{\'\i}pio and Nunes, C{\'e}lia and Jatowt, Adam},
  journal={Information Sciences},
  volume={509},
  pages={257--289},
  year={2020},
  publisher={Elsevier}
}

@inproceedings{Lee2019icomposer,
  title={iComposer: An automatic songwriting system for Chinese popular music},
  author={Lee, Hsin-Pei and Fang, Jhih-Sheng and Ma, Wei-Yun},
  booktitle={Proceedings of the 2019 Conference of the North American Chapter of the Association for Computational Linguistics (Demonstrations)},
  pages={84--88},
  year={2019}
}

@inproceedings{Potash2015Ghostwriter,
  title={Ghostwriter: Using an lstm for automatic rap lyric generation},
  author={Potash, Peter and Romanov, Alexey and Rumshisky, Anna},
  booktitle={Proceedings of the 2015 Conference on Empirical Methods in Natural Language Processing},
  pages={1919--1924},
  year={2015}
}

@article{achiam2023gpt,
  title={Gpt-4 technical report},
  author={Achiam, Josh and Adler, Steven and Agarwal, Sandhini and Ahmad, Lama and Akkaya, Ilge and Aleman, Florencia Leoni and Almeida, Diogo and Altenschmidt, Janko and Altman, Sam and Anadkat, Shyamal and others},
  journal={arXiv preprint arXiv:2303.08774},
  year={2023}
}

@article{ouyang2022training,
  title={Training language models to follow instructions with human feedback},
  author={Ouyang, Long and Wu, Jeffrey and Jiang, Xu and Almeida, Diogo and Wainwright, Carroll and Mishkin, Pamela and Zhang, Chong and Agarwal, Sandhini and Slama, Katarina and Ray, Alex and others},
  journal={Advances in neural information processing systems},
  volume={35},
  pages={27730--27744},
  year={2022}
}

@article{Nick2024Exploring,
  title={Exploring Variational Auto-encoder Architectures, Configurations, and Datasets for Generative Music Explainable AI},
  author={Nick, Bryan-Kinns and Bingyuan, Zhang and Songyan, Zhao and Berker, Banar.},
  journal={Machine Intelligence Research},
  volume={21},
  pages={29-45},
  year={2024},
 DOI = {10.1007/s11633-023-1457-1}
}

@INPROCEEDINGS{Ram2021-hl,
  title     = "Say what? Collaborative pop lyric generation using multitask
               transfer learning",
  author    = "Ram, Naveen and Gummadi, Tanay and Bhethanabotla, Rahul and
               Savery, Richard J and Weinberg, Gil",
  booktitle = "Proceedings of the 9th International Conference on Human-Agent
               Interaction",
  publisher = "ACM",
  address   = "New York, NY, USA",
  month     =  nov,
  year      =  2021,
  language  = "en"
}

@inproceedings{nikolov2020rapformer,
  title={Rapformer: Conditional Rap Lyrics Generation with Denoising Autoencoders},
  author={Nikolov, Nikola I and Malmi, Eric and Northcutt, Curtis and Parisi, Loreto},
  booktitle={Proceedings of the 13th International Conference on Natural Language Generation},
  pages={360--373},
  year={2020}
}

@ARTICLE{Rafailov2023-et,
  title         = "Direct Preference Optimization: Your Language Model is
                   Secretly a Reward Model",
  author        = "Rafailov, Rafael and Sharma, Archit and Mitchell, Eric and
                   Ermon, Stefano and Manning, Christopher D and Finn, Chelsea",
  journal       = "arXiv [cs.LG]",
  month         =  may,
  year          =  2023,
  archivePrefix = "arXiv",
  primaryClass  = "cs.LG"
}

@ARTICLE{Christiano2017-je,
  title         = "Deep reinforcement learning from human preferences",
  author        = "Christiano, Paul and Leike, Jan and Brown, Tom B and Martic,
                   Miljan and Legg, Shane and Amodei, Dario",
  journal       = "arXiv [stat.ML]",
  month         =  jun,
  year          =  2017,
  archivePrefix = "arXiv",
  primaryClass  = "stat.ML"
}

@misc{lyricstudio,
  author = {LyricStudio Team},
  title = {{L}yric{S}tudio --- lyricstudio.net},
  year = {2024},
  url = {https://lyricstudio.net},
  lastaccessed = {September 9, 2024}
}

@book{pamela-aa,
    author = "Pamela Burnard",
    title = "Musical Creativities in Practice",
    publisher = "Oxford University Press",
    year = 2012
}

@book{pat-aa,
    author = "Pat Pattison",
    title = "Writing Better Lyrics (2nd. ed.)",
    publisher = "Writer's Digest Books",
    year = 2010
}

@book{andrea-aa,
    author = "Andrea Stolpe",
    title = "Popular Lyric Writing: 10 Steps to Effective Storytelling",
    publisher = "Berklee Press",
    year = 2007
}

@book{pamela-bb,
    author = "Pamela Phillips Oland",
    title = "The Art of writing Great Lyrics (1st. ed.)",
    publisher = "Allworth",
    year = 2001
}

@book{jason-aa,
    author = "Jason Blume",
    title = "Six Steps to Songwriting success  (2nd. ed.)",
    publisher = "Billboard Books",
    year = 2008
}

@book{Tommy-aa,
    author = "Tommy Swindali",
    title = "Music Elements: Music Theory, Songwriting, Lyrics and Creativity Explained",
    publisher = "Independently published",
    year = 2019
}

@book{Takamitsu-aa,
    author = "Takamitsu Shimazaki",
    title = "Sakushi no Benkyobon [A Study Book For Lyricist]",
    publisher = "Rittor Music (in Japan)",
    year = 2015
}

@book{Himi-aa,
    author = "Himi Izutsu",
    title = "Zero kara no Sakushi Nyuumon [Introduction to Lyric Writing from Scratch]",
    publisher = "Yamaha Music Entertainment Holdings, Inc.",
    year = 2015
}

@book{Tatsuji-aa,
    author = "Tatsuji Ueda",
    title = "Yoku wakaru Sakushi no Kyokasho [Textbook of Lyric Writing]",
    publisher = "Yamaha Music Entertainment Holdings, Inc.",
    year = 2010
}

@book{Oscar-aa,
    author = "Oscar Hammerstein",
    title = "Lyrics by Oscar Hammerstein II",
    publisher = "Hal Leonard Publishing",
    year = 1985
}

@book{Sheppard2009-aa,
  title={On Some Faraway Beach: The Life and Times of Brian Eno},
  author={Sheppard, David},
  year={2009},
  publisher={Chicago Review Press}
}

@book{Zollo-aa,
    author = "Paul Zollo (Revised ed.)",
    title = "Songwriters On Songwriting",
    publisher = "Da Capo Press",
    year = 2003
}

@ARTICLE{Negus2015-bd,
  title     = "Songwriters and song lyrics: architecture, ambiguity and
               repetition",
  author    = "Negus, Keith and Astor, Pete",
  journal   = "Pop. Music",
  publisher = "Cambridge University Press (CUP)",
  volume    =  34,
  number    =  2,
  pages     = "226--244",
  month     =  may,
  year      =  2015,
  language  = "en"
}

@MASTERSTHESIS{Kardzha2024-vx,
  title   = "Intuition and intention in the songwriting process of contemporary
             lyric-driven songs",
  author  = "Kardzha, Deniz",
  address = "Tampere, Finland",
  year    =  2024,
  school  = "Tampere University of Applied Sciences"
}

@ARTICLE{Reinhert2019-ue,
  title     = "Lyric approaches to songwriting in the classroom",
  author    = "Reinhert, Kat",
  journal   = "J. Popul. Music Educ.",
  publisher = "Intellect",
  volume    =  3,
  number    =  1,
  pages     = "129--139",
  month     =  apr,
  year      =  2019,
  language  = "en"
}

@ARTICLE{Dalton2006-fi,
  title     = "The grief song-writing process with bereaved adolescents: An
               integrated grief model and music therapy protocol",
  author    = "Dalton, T A and Krout, R E",
  journal   = "Music Ther. Perspect.",
  publisher = "Oxford University Press (OUP)",
  volume    =  24,
  number    =  2,
  pages     = "94--107",
  month     =  jan,
  year      =  2006,
  language  = "en"
}

@INPROCEEDINGS{Oura2010-fl,
  title     = "Recent development of the {HMM}-based singing voice synthesis
               system — Sinsy",
  author    = "Oura, Keiichiro and Mase, Ayami and Yamada, Tomohiko and Muto,
               Satoru and Nankaku, Yoshihiko and Tokuda, Keiichi",
  booktitle = "Proc. SSW 2010",
  pages     = "211--216",
  year      =  2010
}

@inproceedings{Larsen-ab,
author = {Gammelg\r{a}rd-Larsen, Anders and van Berkel, Niels and Skov, Mikael B. and Kjeldskov, Jesper},
title = {Designing for Human-AI Interaction: Comparing Intermittent, Continuous, and Proactive Interactions for a Music Application},
year = {2024},
isbn = {9798400703317},
publisher = {Association for Computing Machinery},
address = {New York, NY, USA},
url = {https://doi.org/10.1145/3613905.3650886},
doi = {10.1145/3613905.3650886},
booktitle = {Extended Abstracts of the CHI Conference on Human Factors in Computing Systems},
articleno = {105},
numpages = {8},
location = {Honolulu, HI, USA},
series = {CHI EA '24}
}

@INPROCEEDINGS{Padmakumar2023-dg,
  title     = "Does Writing with Language Models Reduce Content Diversity?",
  author    = "Padmakumar, Vishakh and He, He",
  booktitle = "The Twelfth International Conference on Learning Representations",
  month     =  oct,
  year      =  2023
}

@INPROCEEDINGS{Malmi2016-dl,
  title     = "{DopeLearning}: A computational approach to rap lyrics generation",
  author    = "Malmi, Eric and Takala, Pyry and Toivonen, Hannu and Raiko,
               Tapani and Gionis, Aristides",
  booktitle = "Proceedings of the 22nd ACM SIGKDD International Conference on
               Knowledge Discovery and Data Mining",
  publisher = "ACM",
  address   = "New York, NY, USA",
  month     =  aug,
  year      =  2016,
  language  = "en"
}

@INCOLLECTION{Vechtomova2023-pe,
  title     = "{LyricJam} sonic: A generative system for real-time composition
               and musical improvisation",
  author    = "Vechtomova, Olga and Sahu, Gaurav",
  booktitle = "Artificial Intelligence in Music, Sound, Art and Design",
  publisher = "Springer Nature Switzerland",
  address   = "Cham",
  pages     = "292--307",
  series    = "Lecture notes in computer science",
  year      =  2023,
  language  = "en"
}

@ARTICLE{Vechtomova2021-ld,
  title         = "{LyricJam}: A system for generating lyrics for live
                   instrumental music",
  author        = "Vechtomova, Olga and Sahu, Gaurav and Kumar, Dhruv",
  journal       = "arXiv [cs.SD]",
  month         =  jun,
  year          =  2021,
  archivePrefix = "arXiv",
  primaryClass  = "cs.SD"
}

@misc{Dodson2024,
  author = {Marty Dodson},
  title = {The Single Biggest Distinction Between Pro And Amateur Songwriters},
  year = {2024},
  url = {https://songtown.com/on-songwriting/pro-vs-amateur-songwriters/},
  note = {Accessed: 2025-04-07}
}

@book{baskerville2013music,
  title     = {Music Business Handbook and Career Guide},
  author    = {Baskerville, David and Baskerville, Tim},
  edition   = {10th},
  year      = {2013},
  publisher = {SAGE Publications},
  isbn      = {9781452242200}
}

@INPROCEEDINGS{Karolus2023-ia,
  title     = "Your text is hard to read: Facilitating readability awareness to
               support writing proficiency in text production",
  author    = "Karolus, Jakob and Feger, Sebastian S and Schmidt, Albrecht and
               Woźniak, Paweł W",
  booktitle = "Proceedings of the 2023 ACM Designing Interactive Systems
               Conference",
  publisher = "ACM",
  address   = "New York, NY, USA",
  pages     = "147--160",
  month     =  jul,
  year      =  2023,
  language  = "en"
}

@INPROCEEDINGS{Laban2024-ya,
  title     = "Beyond the chat: Executable and verifiable text-editing with
               {LLMs}",
  author    = "Laban, Philippe and Vig, Jesse and Hearst, Marti and Xiong,
               Caiming and Wu, Chien-Sheng",
  booktitle = "Proceedings of the 37th Annual ACM Symposium on User Interface
               Software and Technology",
  publisher = "ACM",
  address   = "New York, NY, USA",
  volume    =  10,
  pages     = "1--23",
  month     =  oct,
  year      =  2024,
  language  = "en"
}

@INPROCEEDINGS{Afrin2021-up,
  title     = "Effective interfaces for student-driven revision sessions for
               argumentative writing",
  author    = "Afrin, Tazin and Kashefi, Omid and Olshefski, Christopher and
               Litman, Diane and Hwa, Rebecca and Godley, Amanda",
  booktitle = "Proceedings of the 2021 CHI Conference on Human Factors in
               Computing Systems",
  publisher = "ACM",
  address   = "New York, NY, USA",
  month     =  may,
  year      =  2021,
  language  = "en"
}

@INPROCEEDINGS{Wambsganss2022-ym,
  title     = "Adaptive empathy learning support in peer review scenarios",
  author    = "Wambsganss, Thiemo and Soellner, Matthias and Koedinger, Kenneth
               R and Leimeister, Jan Marco",
  booktitle = "CHI Conference on Human Factors in Computing Systems",
  publisher = "ACM",
  address   = "New York, NY, USA",
  month     =  apr,
  year      =  2022,
  language  = "en"
}

@inproceedings{
Zhang2024-nq,
title={Adaptable Logical Control for Large Language Models},
author={Honghua Zhang and Po-Nien Kung and Masahiro Yoshida and Guy Van den Broeck and Nanyun Peng},
booktitle={The Thirty-eighth Annual Conference on Neural Information Processing Systems},
year={2024},
url={https://openreview.net/forum?id=58X9v92zRd}
}

@ARTICLE{Li2024-kr,
  title         = "Personalized Language Modeling from Personalized Human
                   Feedback",
  author        = "Li, Xinyu and Lipton, Zachary C and Leqi, Liu",
  journal       = "arXiv [cs.CL]",
  month         =  feb,
  year          =  2024,
  archivePrefix = "arXiv",
  primaryClass  = "cs.CL"
}

@ARTICLE{Liu2025-sz,
  title         = "A survey of personalized Large Language Models: Progress and
                   future directions",
  author        = "Liu, Jiahong and Qiu, Zexuan and Li, Zhongyang and Dai,
                   Quanyu and Zhu, Jieming and Hu, Minda and Yang, Menglin and
                   King, Irwin",
  journal       = "arXiv [cs.AI]",
  month         =  feb,
  year          =  2025,
  archivePrefix = "arXiv",
  primaryClass  = "cs.AI"
}

@INPROCEEDINGS{Biermann2022-jk,
  title     = "From tool to companion: Storywriters want {AI} writers to respect
               their personal values and writing strategies",
  author    = "Biermann, Oloff C and Ma, Ning F and Yoon, Dongwook",
  booktitle = "Designing Interactive Systems Conference",
  publisher = "ACM",
  address   = "New York, NY, USA",
  month     =  jun,
  year      =  2022
}

@ARTICLE{Haase2024-ql,
  title         = "Human-{AI} co-creativity: Exploring synergies across levels
                   of creative collaboration",
  author        = "Haase, Jennifer and Pokutta, Sebastian",
  journal       = "arXiv [cs.HC]",
  month         =  nov,
  year          =  2024,
  archivePrefix = "arXiv",
  primaryClass  = "cs.HC"
}

@INPROCEEDINGS{Xu2024-dj,
  title     = "What makes it mine? Exploring psychological ownership over
               human-{AI} co-creations",
  author    = "Xu, Yuxin and Cheng, Mengqiu and Kuzminykh, Anastasia",
  booktitle = "Graphics Interface",
  publisher = "ACM",
  address   = "New York, NY, USA",
  pages     = "1--8",
  month     =  jun,
  year      =  2024,
  language  = "en"
}

@INPROCEEDINGS{Weber2024-ek,
  title     = "{LegalWriter}: An intelligent writing support system for
               structured and persuasive legal case writing for novice law
               students",
  author    = "Weber, Florian and Wambsganss, Thiemo and Neshaei, Seyed Parsa
               and Soellner, Matthias",
  booktitle = "Proceedings of the CHI Conference on Human Factors in Computing
               Systems",
  publisher = "ACM",
  address   = "New York, NY, USA",
  volume    =  12,
  pages     = "1--23",
  month     =  may,
  year      =  2024,
  language  = "en"
}

@INPROCEEDINGS{Dhillon2024-sx,
  title     = "Shaping Human-{AI} Collaboration: Varied Scaffolding Levels in
               Co-writing with Language Models",
  author    = "Dhillon, Paramveer S and Molaei, Somayeh and Li, Jiaqi and Golub,
               Maximilian and Zheng, Shaochun and Robert, Lionel Peter",
  booktitle = "Proceedings of the CHI Conference on Human Factors in Computing
               Systems",
  publisher = "Association for Computing Machinery",
  address   = "New York, NY, USA",
  number    = "Article 1044",
  pages     = "1--18",
  series    = "CHI '24",
  month     =  may,
  year      =  2024
}

@INPROCEEDINGS{Reza2024-os,
  title     = "{ABScribe}: Rapid exploration \& organization of multiple writing
               variations in human-{AI} co-writing tasks using large language
               models",
  author    = "Reza, Mohi and Laundry, Nathan M and Musabirov, Ilya and
               Dushniku, Peter and Yu, Zhi Yuan “michael” and Mittal, Kashish
               and Grossman, Tovi and Liut, Michael and Kuzminykh, Anastasia and
               Williams, Joseph Jay",
  booktitle = "Proceedings of the CHI Conference on Human Factors in Computing
               Systems",
  publisher = "ACM",
  address   = "New York, NY, USA",
  volume    =  15,
  pages     = "1--18",
  month     =  may,
  year      =  2024,
  language  = "en"
}

@INPROCEEDINGS{Benharrak2024-ov,
  title     = "Writer-Defined {AI} Personas for On-Demand Feedback Generation",
  author    = "Benharrak, Karim and Zindulka, Tim and Lehmann, Florian and
               Heuer, Hendrik and Buschek, Daniel",
  booktitle = "Proceedings of the CHI Conference on Human Factors in Computing
               Systems",
  publisher = "Association for Computing Machinery",
  address   = "New York, NY, USA",
  number    = "Article 1049",
  pages     = "1--18",
  series    = "CHI '24",
  month     =  may,
  year      =  2024
}

@INPROCEEDINGS{Zhou2021-cb,
  title     = "Interactive Exploration-Exploitation Balancing for Generative
               Melody Composition",
  author    = "Zhou, Yijun and Koyama, Yuki and Goto, Masataka and Igarashi,
               Takeo",
  booktitle = "26th International Conference on Intelligent User Interfaces",
  publisher = "Association for Computing Machinery",
  address   = "New York, NY, USA",
  pages     = "43--47",
  series    = "IUI '21",
  month     =  apr,
  year      =  2021
}

@INPROCEEDINGS{Koyama2022-kb,
  title     = "{BO} as Assistant: Using Bayesian Optimization for Asynchronously
               Generating Design Suggestions",
  author    = "Koyama, Yuki and Goto, Masataka",
  booktitle = "Proceedings of the 35th Annual ACM Symposium on User Interface
               Software and Technology",
  publisher = "Association for Computing Machinery",
  address   = "New York, NY, USA",
  number    = "Article 77",
  pages     = "1--14",
  series    = "UIST '22",
  month     =  oct,
  year      =  2022
}

@INPROCEEDINGS{Frid2020-fz,
  title     = "Music creation by example",
  author    = "Frid, Emma and Gomes, Celso and Jin, Zeyu",
  booktitle = "Proceedings of the 2020 CHI Conference on Human Factors in
               Computing Systems",
  publisher = "ACM",
  address   = "New York, NY, USA",
  month     =  apr,
  year      =  2020,
  language  = "en"
}

@INPROCEEDINGS{Kim2024-ou,
  title     = "{DiaryMate}: Understanding user perceptions and experience in
               human-{AI} collaboration for personal journaling",
  author    = "Kim, Taewan and Shin, Donghoon and Kim, Young-Ho and Hong,
               Hwajung",
  booktitle = "Proceedings of the CHI Conference on Human Factors in Computing
               Systems",
  publisher = "ACM",
  address   = "New York, NY, USA",
  pages     = "1--15",
  month     =  may,
  year      =  2024,
  language  = "en"
}

@INPROCEEDINGS{Louie2020-um,
  title     = "Novice-{AI} music co-creation via {AI}-steering tools for deep
               generative models",
  author    = "Louie, Ryan and Coenen, Andy and Huang, Cheng Zhi and Terry,
               Michael and Cai, Carrie J",
  booktitle = "Proceedings of the 2020 CHI Conference on Human Factors in
               Computing Systems",
  publisher = "ACM",
  address   = "New York, NY, USA",
  month     =  apr,
  year      =  2020,
  language  = "en"
}

@INPROCEEDINGS{Jakesch2023-ih,
  title     = "Co-Writing with Opinionated Language Models Affects Users’ Views",
  author    = "Jakesch, Maurice and Bhat, Advait and Buschek, Daniel and
               Zalmanson, Lior and Naaman, Mor",
  booktitle = "Proceedings of the 2023 CHI Conference on Human Factors in
               Computing Systems",
  publisher = "Association for Computing Machinery",
  address   = "New York, NY, USA",
  number    = "Article 111",
  pages     = "1--15",
  series    = "CHI '23",
  month     =  apr,
  year      =  2023
}

@INPROCEEDINGS{Dang2022-te,
  title     = "Beyond Text Generation: Supporting Writers with Continuous
               Automatic Text Summaries",
  author    = "Dang, Hai and Benharrak, Karim and Lehmann, Florian and Buschek,
               Daniel",
  booktitle = "Proceedings of the 35th Annual ACM Symposium on User Interface
               Software and Technology",
  publisher = "Association for Computing Machinery",
  address   = "New York, NY, USA",
  number    = "Article 98",
  pages     = "1--13",
  series    = "UIST '22",
  month     =  oct,
  year      =  2022
}

@INPROCEEDINGS{Kobayashi2024-jx,
  title     = "Novice-centered application design for music creation",
  author    = "Kobayashi, Atsuya and Sato, Tetsuro and Tateno, Kei",
  booktitle = "Designing Interactive Systems Conference",
  publisher = "ACM",
  address   = "New York, NY, USA",
  month     =  jul,
  year      =  2024
}

@INPROCEEDINGS{Zhao2024-ic,
  title     = "{REFFLY}: Melody-constrained lyrics editing model",
  author    = "Zhao, Songyan and Li, Bingxuan and Tian, Yufei and Peng, Nanyun",
  booktitle = "Proceedings of the 2025 Conference of the Nations of the Americas
               Chapter of the Association for Computational Linguistics: Human
               Language Technologies (Volume 1: Long Papers)",
  publisher = "Association for Computational Linguistics",
  address   = "Stroudsburg, PA, USA",
  pages     = "11295--11315",
  year      =  2025
}

@INPROCEEDINGS{Fu2025-ui,
  title     = "Exploring the collaborative co-creation process with {AI}: A case
               study in novice music production",
  author    = "Fu, Yue and Newman, Michele and Going, Lewis and Feng, Qiuzi and
               Lee, Jin Ha",
  booktitle = "Proceedings of the 2025 ACM Designing Interactive Systems
               Conference",
  publisher = "ACM",
  address   = "New York, NY, USA",
  pages     = "1298--1312",
  month     =  jul,
  year      =  2025,
  language  = "en"
}

@ARTICLE{Chae2025-gh,
  title         = "Song form-aware full-song text-to-lyrics generation with
                   multi-level granularity syllable count control",
  author        = "Chae, Yunkee and Shin, Eunsik and Hwang, Suntae and Paik,
                   Seungryeol and Lee, Kyogu",
  journal       = "arXiv [cs.CL]",
  month         =  jun,
  year          =  2025,
  archivePrefix = "arXiv",
  primaryClass  = "cs.CL"
}

@INPROCEEDINGS{Morris2025-ki,
  title     = "Expanding the {HAISP} Dataset: {AI}'s Impact on Songwriting
               Across Two Contests",
  author    = "Morris, Lidia and Newman, Michele and Tang, Xinya and Singh,
               Renee and Vásquez, Marcel Vélez and Leger, Rebecca and Lee, Jin
               Ha",
  booktitle = "Proceedings of the 26th International Society for Music
               Information Retrieval Conference, ISMIR 2025",
  year      =  2025
}

@INPROCEEDINGS{Kim2025-xf,
  title     = "Amuse: Human-{AI} collaborative songwriting with multimodal
               inspirations",
  author    = "Kim, Yewon and Lee, Sung-Ju and Donahue, Chris",
  booktitle = "Proceedings of the 2025 CHI Conference on Human Factors in
               Computing Systems",
  publisher = "ACM",
  address   = "New York, NY, USA",
  pages     = "1--28",
  month     =  apr,
  year      =  2025,
  language  = "en"
}

@INPROCEEDINGS{Zhang2024-fz,
  title     = "Syllable-level lyrics generation from melody exploiting
               character-level language model",
  author    = "Zhang, Zhe and Lasocki, Karol and Yu, Yi and Takasu, Atsuhiro",
  booktitle = "Findings of the Association for Computational Linguistics: EACL
               2024",
  pages     = "1336--1346",
  year      =  2024
}

\clearpage
\appendix

\section{Interview Protocol for the Formative Study}
\label{app:interview_protocol}

This appendix presents the interview questions used in the formative study. 
The semi-structured interviews were organized into the following four sections.
\paragraph{1. Background}

\begin{itemize}
  \item Age, major (degree), or current occupation.
  \item Do you have experience playing musical instruments? If so, when did you start?
  \item Approximately how many songs have you written lyrics for or composed in the past? When did you create them?
  \item For what purposes do you usually engage in songwriting? 
  (e.g., uploading songs to social media individually, participating in a band, commissioned work)
\end{itemize}

\paragraph{2. Songwriting Workflow}

\begin{itemize}
  \item Please describe your most recent songwriting experience.
  \item Can you describe your overall workflow during this songwriting process?
  \item Is this workflow generally the same as other songwriting experiences? If there are different workflows, please describe them.
  \item During this process, what were your specific requirements or inputs? 
  What formats or constraints did these requirements have?
  \item What tools did you use to complete this songwriting task?
  \item In your opinion, which aspects of the songwriting process could be improved to increase efficiency?
\end{itemize}

\paragraph{3. Songwriting Needs and Challenges}

\begin{itemize}
  \item What do you think are the most important aspects of lyric writing?
  \item What were the main difficulties or challenges you encountered during the lyric-writing process?
  \item In the songwriting process, how do you or your clients evaluate whether lyrics are good?
  Are there any specific criteria or standards?
\end{itemize}

\paragraph{4. Use of AI Songwriting Tools}

\begin{itemize}
  \item Have you used AI-based songwriting or lyric generation tools? 
  If so, please describe your experiences.
  \item What do you think are the main problems or limitations of current AI songwriting tools?
  \item If there were a tool that could better support your songwriting process, 
  what aspects would you hope it could assist with?
  \item During the songwriting process, which parts would you prefer to keep under human control, 
  and which parts would you be comfortable delegating to AI systems?
\end{itemize}

\section{Melody-fitting Model Implementation Detail}
\label{app:implementation}
Reffly is trained exclusively on lyrics data, adhering to synthesized constraints during training and refining the draft to align with real constraints extracted from the given melody during inference. This design ensures the revised lyrics is singable and has good prosody. We re-implement Reffly using GPT \cite{openai_gpt3_5} as a pre-trained model, instead of the original LLaMA2-13b model. To enhance lyric revision capabilities, we fine-tuned GPT-4o-mini, optimizing for quicker response times, low latency, and stability \cite{ouyang2022training, achiam2023gpt}. We  prepared a dataset with 20,000 song-lyrics collected from the internet to instruct the model to generate original lyrics by revising draft lyrics, following music constraints. Following the setup of REFFLY, the input of dataset are rephrased input, title, music constraint, previously generated lyrics, and assembled instruction. The output is the revised lyrics, and some explanation. During training, the revision model is guided by pseudo melody constraints derived from the original lyrics, enabling it to follow real melody constraints during inference. Using this dataset, we then proceeded to fine-tune the GPT4-o-mini model for two epochs. We improved the framework's efficiency by implementing parallel revision, enabling simultaneous multi-line lyric modifications. 


To increase the chances of users finding inspiring lyrics, the system presents four lyric options at once. To enhance the diversity of these options, the original draft is rephrased four times, and these variations are input into Reffly. Additionally, we designed a quality control function with YAKE \cite{Ricardo2020Yake}  to ensure high-caliber, instruction-aligned lyric revisions. This process involves ranking the revised lyrics based on the number of satisfied kewords/ key phrases, with the top four selections being presented as output.
Each output option is evaluated using Reffly's modules, which score the alignment of syllables and prosody with the melody, and the options are ordered according to these scores. Specifically, all candidates are ranked by the number of prominent lyrical words mapped to prominent musical notes—the more prominent words that align with the notes, the higher the score.

\section{Prompt Templates}
\label{app:prompts}

In this section, we provide the few-shot prompt templates fore each features in the \projectname.

\subsection{Draft Generation}

\begin{lstlisting}
[INSTRUCTIONS]
Task: Compose song lyrics according to the specified parameters.
General Instructions:
1. Adhere strictly to the given theme and title (if provided).
2. Incorporate key phrases as directed.
3. Maintain the specified number of lines.
4. Return only the lyrics, with each line as a separate string.
5. Do not include the title, theme, or any additional commentary.

[EXAMPLE 1]
Input:
- Title: "Temporal Odyssey"
- Theme: The passage of time
- Key phrases: Fleeting moments, Eternal change
- Lines: 4
Output:
Fleeting moments slip through hourglass sands
Centuries unfold in the blink of an eye
Eternal change shapes our mortal plans
Time's river flows, never asking why

[YOUR TASK]
Please compose lyrics based on the following specifications:
- Title: {*title*}
- Theme: {*theme*}
- Key phrases: {*key_words*}
- Lines: {*num_of_lines*}
\end{lstlisting}

\subsection{AI-assisted Brainstorm}

\begin{lstlisting}

[INSTRUCTIONS]
You are a song writer tasked with generating phrases for songs based on given inputs. 
For each example, provide 5 relevant phrases that could be used in the song. 
Return your response as a list of strings. No additional explanation is needed.

[EXAMPLE 1]
Input: love at first sight
Theme: romance
Song Title: "Cupid's Arrow"
Output: [Love struck me like lightning, 
Your eyes told a story, 
Hearts beating in sync, 
A moment that changed everything, Destiny's perfect timing]

[EXAMPLE 2]
Input: environmental conservation
Theme: nature
Song Title: "Green Earth"
Output: [Forests breathing life, 
Oceans cry for healing, 
Every action counts, 
Nurture Mother Nature, Together we can save our home]

[EXAMPLE 3]
Input: overcoming adversity
Theme: personal growth
Song Title: "Rise Above"
Output: [Stronger with each fall, 
Breaking through barriers, 
The fire within ignites, 
Victory over fear, Embracing the journey]

[YOUR TASK]
You are the song writer writing a song called "{*title*}",
with the theme of {*theme*}. 
Generate 5 relevant phrases of provided input. 
Return in the string of list. 
No verbose output. 
Example output format: [phrase1, phrase2, phrase3, phrase4, phrase5]. 
Here is the input: {*input*}
\end{lstlisting}

\subsection{Rhyming Recommendation}

\begin{lstlisting}
[INSTRUCTIONS]
Please provide rhyming words based on the following criteria. 
For each example, give 8 words that rhyme with the given word. 
Follow any additional instructions about syllable count or theme.
Only return words in a string, separated by commas. No additional explanation needed.

[EXAMPLE 1]
Input: cat, no syllable restriction, animal theme
Output: bat, rat, gnat, mat, sat, flat, spat, hat

[EXAMPLE 2]
Input: light, 2 syllables, no theme
Output: invite, ignite, excite, delight, tonight, unite, alight, despite

[EXAMPLE 3]
Input: education, 3 syllables, school theme
Output: graduation, vocation, formation, notation, vacation, equation, rotation, location

[YOUR TASK]
Please give me 8 words {*syllable_condition}* that ends 
with same rhyme of the word '{word}'. 
{theme_condition} Only return words in a string, 
separate by comma. No verbose output.
\end{lstlisting}

\clearpage
\onecolumn

\subsection{Tool Overview}
\label{app:tool_overview}
\begin{figure}[H]  

    \centering
    \includegraphics[width=1\linewidth]{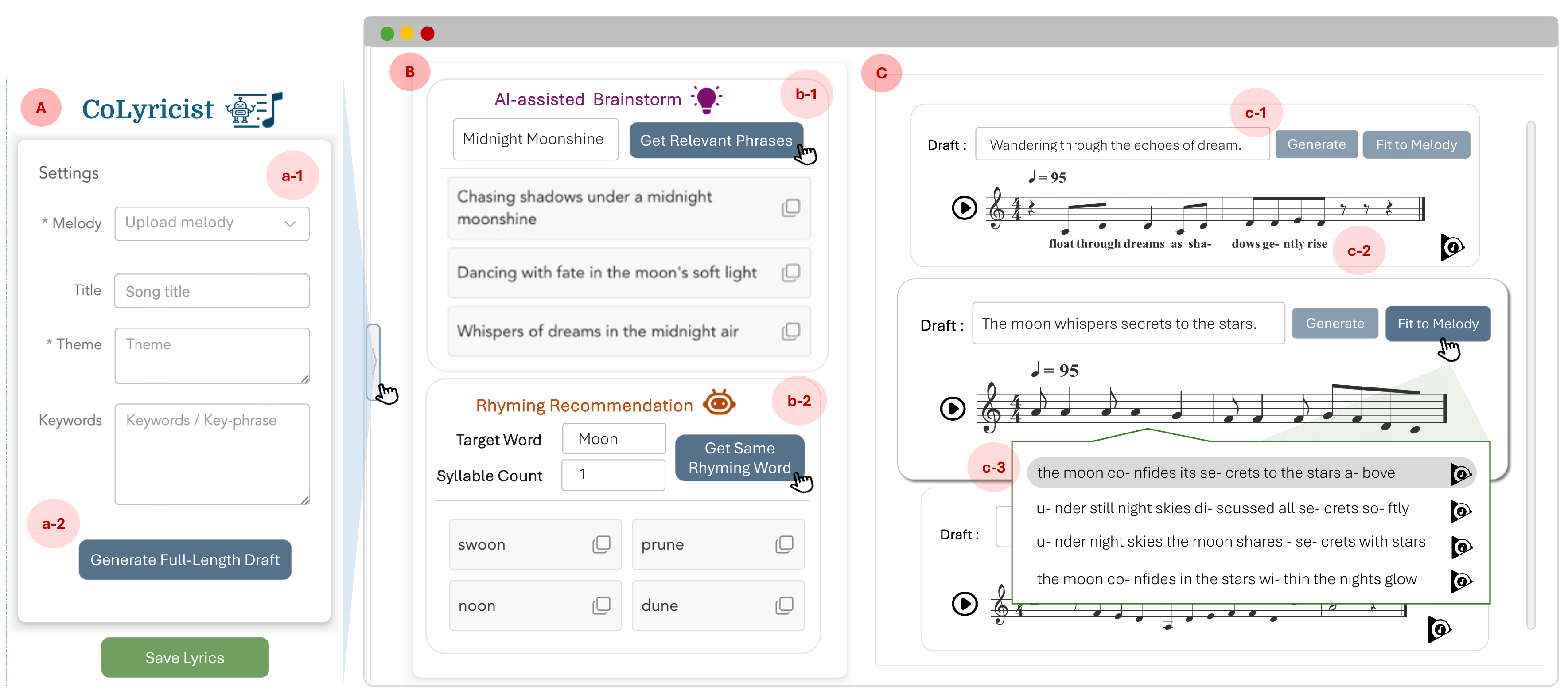}
    \caption{The user interface of CoLyricist includes: (A) Settings panel for configuring the song's melody, theme, and other parameters; (B) Ideation panel for assisting in brainstorming and providing rhyming recommendations; (C) Editing panel for drafting and refining lyrics, where AI provides automated revised lyrics that fit with the melody.}
    \Description{Annotated overview of the CoLyricist user interface divided into three main panels. Panel A on the left is the Settings panel, where users configure core inputs such as melody upload, song title, theme, and optional keywords, and can generate a full-length draft based on these inputs. Panel B in the center is the Ideation panel, which includes an AI-assisted Brainstorm feature for generating thematically related phrases from a given keyword or phrase, and a Rhyming Recommendation feature that provides rhyming word suggestions based on a target word and specified syllable count. Panel C on the right is the Editing panel, where users draft lyrics line by line, view the corresponding musical staff, listen to melody playback, and request AI-assisted revisions that fit the melody. Highlighted sublabels and callouts indicate specific interactive elements, such as generation buttons, melody fitting actions, and lists of revised lyric candidates, illustrating how users move between configuration, ideation, and iterative lyric refinement.}
    \label{fig:tool_overview}
\end{figure}

\clearpage

\subsection{Selected Lyrics created by users}
\label{app:lyrics}

\begin{figure*}[h]
    \centering
    \includegraphics[width=1\linewidth]{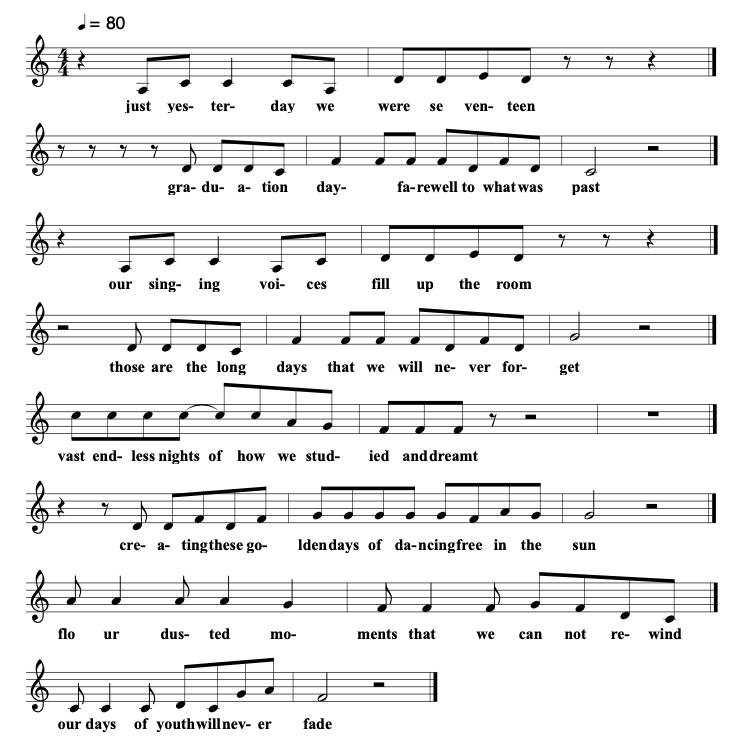}
    \caption{Selected Lyrics 1: Created by novice participants}
    \Description{Musical score excerpt with lyrics showing a song created by a novice participant. The figure presents staff notation with a tempo marking at the top, followed by multiple measures of melody with lyrics aligned beneath the notes. Each line displays syllables segmented across notes to match the melodic rhythm, illustrating how the generated lyrics are mapped onto the melody throughout the song. This example serves as a representative output of lyrics created by novice users using the CoLyricist system.}
    \label{fig:lyrics_1}
\end{figure*}

\begin{figure*}[h]
    \centering
    \includegraphics[width=1\linewidth]{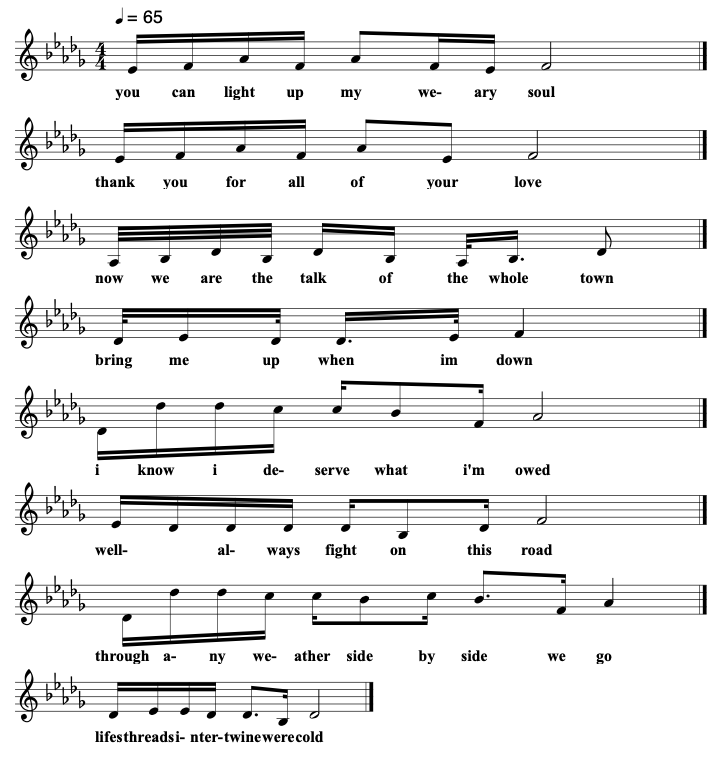}
    \caption{Selected Lyrics 2: Created by experienced participants}
    \Description{Musical score excerpt with lyrics showing a song created by a experienced participant. The figure presents staff notation with a tempo marking at the top, followed by multiple measures of melody with lyrics aligned beneath the notes. Each line displays syllables segmented across notes to match the melodic rhythm, illustrating how the generated lyrics are mapped onto the melody throughout the song. This example serves as a representative output of lyrics created by experienced users using the CoLyricist system.}

    \label{fig:lyrics_2}
\end{figure*}

\end{document}